\documentclass[aps,prl,reprint,groupedaddress,showpacs,showkeys,amsmath,amssymb,]{revtex4-1}
\usepackage{graphicx}
\usepackage{upgreek}

\begin{document}


\title{Raman sideband cooling  of  {$^{25}$Mg$^+$} -{$^{27}$Al$^+$} ions pair and observation of quantum logic spectra}

\author{Hong-Li Liu}
\author{Ze-Tian Xu}
\author{Zhi-Yu Ma} 
\author{ Wen-Zhe Wei}
\author{Hao Peng}
\author{ Wen-Hao Yuan}
\author{Huang Che} 
\author{ Fei-Hu Cheng}
\author{ Zhi-Yuan Wang}
\author{Ke Deng}
\email {ke.deng@hust.edu.cn}
\author{Jie Zhang}
\author{Ze-Huang Lu}
\email{zehuanglu@hust.edu.cn}

\affiliation{MOE Key Laboratory of Fundamental Physical Quantities Measurement $\&$ Hubei Key Laboratory of Gravitation and Quantum Physics, PGMF and School of Physics, Huazhong University of Science and Technology, Wuhan 430074,  P. R. China.}

\date{\today}

\begin{abstract}
We perform quantum logic spectroscopy (QLS) on {$^{27}$Al$^+$} ion $^1$S$_0$  - $^3$P$_1$ transition, which is an important  step toward the QLS based {$^{27}$Al$^+$} ion optical clock. As a precondition of QLS, both the stretch (STR) mode and the common (COM) mode  of the {$^{27}$Al$^+$} and {$^{25}$Mg$^+$} ions pair are cooled to the vibrational ground state by Raman sideband cooling.  The mean phonon number  is measured to be 0.10(1) for the  STR mode and 0.01(1) for the COM mode, respectively. The  heating rate is evaluated to be  13(3) phonons/s for the STR mode and 5.5(1) phonons/s for the COM mode.
\end{abstract}

\maketitle


Atomic clocks play important roles  both in fundamental science and everyday life \cite{Haensch2006, Chou2010, Steinmetz2008}. Given the increasing demands on better frequency standards for diverse applications in measuring  time variations of fundamental constants \cite{Godun2014,Huntemann2014},
 exploring geophysical phenomena \cite{Bondarescu2015},
detecting gravitational waves \cite{Kolkowitz2016},
testing of general relativity \cite{Delva2017},
and searching for dark matter \cite{Roberts2017},
more stable and more accurate atomic clocks are needed.   In comparison with microwave atomic clocks, optical clocks potentially can have better stability and systematic uncertainty due to higher Q values. Optical clocks has seen rapid  progresses in the past decades. The state-of-the-arts optical clocks have achieved systematic uncertainties at  $10^{-19}$ level \cite{Campbell2017, McGrew2018,Keller2019,Brewer2019},
while the  best Cesium fountain clocks have fractional frequency uncertainties of $1.1 \times 10^{−16}$ \cite{Heavner2014}.

There are several possible routes to develop an optical clock. One is based on trapped ions in a Paul trap, and one is based on
neutral atoms trapped in an optical lattice. Another possibility is to use active optical clocks \cite{Zhuang2014}. Potential choices of single ion species include Hg$^+$ \cite{Oskay2006,Liu2019}, Sr$^+$ \cite{Barwood2014, Dube2014}, Yb$^+$ \cite{King2012, Huntemann2012}, Ca$^+$ \cite{Hashimoto2011, Liu2014}, In$^+$ \cite{Wang2007}, and Ba$^+$ \cite{Koerber2002}, while potential choices of optical lattice clocks and active optical clocks include Sr \cite{Takamoto2011, Bloom2014,Lin2015,Wang2018}, Yb \cite{Hinkley2013,Zhang2016}, and Hg \cite{McFerran2012,Liu2013}.

Due to the  narrow natural linewidth of 8 mHz,  the ${^1S_0} \to {^3P_0}$ transition of {$^{27}$Al$^+$}  at 267.4 nm has  been recognized as an excellent optical clock transition candidate. Its electric quadrupole shift is negligible (J=0), therefore the transition frequency is  insensitive to the electric field gradient in the ion trap \cite{Yu1992}. Furthermore, blackbody radiation contribution to the fractional frequency uncertainty of the {$^{27}$Al$^+$} ion clock is as low as $4 \times 10^{-19}$ at room temperature (300 K), which is the smallest among all atomic species currently under consideration for optical clocks \cite{Safronova2011}. However,  {$^{27}$Al$^+$}  ions could not be cooled directly due to lack of the direct cooling laser at 167 nm. Recently, a two-photon transition cooling based on the transition of $3s^{2}$ $^1S_0 \to 3s3p  {^1D_2} $  at 234 nm laser has been proposed \cite{Zhang2017}, which is still under development.

 To overcome the cooling problem, a method of quantum logic spectroscopy (QLS) was proposed to cool the aluminum ion and to detect the clock transition \cite{Schmidt2005}. 
 Instead of observing the quantum states of  {$^{27}$Al$^+$} ion directly, an auxiliary ion  (for example, Be$^+$, Mg$^+$, Ca$^+$), named as logic ion, is used to coherently map the internal state of the {$^{27}$Al$^+$} ion to logic ion's internal state via the common phonon mode. 
The logic ion provides sympathetic cooling, state initialization, and state detection for the simultaneously trapped {$^{27}$Al$^+$} ion. After Doppler cooling and resolved sideband cooling, both the logic ion and the {$^{27}$Al$^+$} ion can be cooled to the vibrational ground state. The Coulomb interaction of the ions couples their motions. The state detection is achieved through a coherent transfer of the {$^{27}$Al$^+$}  ion's  internal quantum state onto the logic ion. Using QLS method, the first {$^{27}$Al$^+$} ion optical clock was successfully demonstrated at the National Institute of Standards and Technology (NIST) \cite{Rosenband2007}. Since then, several groups are working on QLS based {$^{27}$Al$^+$} ion optical clolcks \cite{Guggemos2015,Cui2018,Ksenia2018,Hannig2019}. In our experiment {$^{25}$Mg$^+$ has been chosen as the logic ion. 

In this Letter, we report the Raman sideband cooling of the  {$^{25}$Mg$^+$} and {$^{27}$Al$^+$} ions pair, the motional mode of the ions pair are cooled to the vibrational ground state with very small average phonon numbers so that the internal states of {$^{25}$Mg$^+$} and {$^{27}$Al$^+$} ions pair could be transferred coherently. The blue-sideband (BSB) of {$^{27}$Al$^+$}$^1$S$_0 {\mid5/2, 5/2>}$ - $^3$P$_1{\mid7/2, 7/2>} $ transition is observed using the QLS technique. The carrier transition is also demonstrated, paving the way for the detection of the $^1$S$_0$  - $^3$P$_0$ clock transition.
 
  \begin{figure}
 \includegraphics[width=8.5 cm]{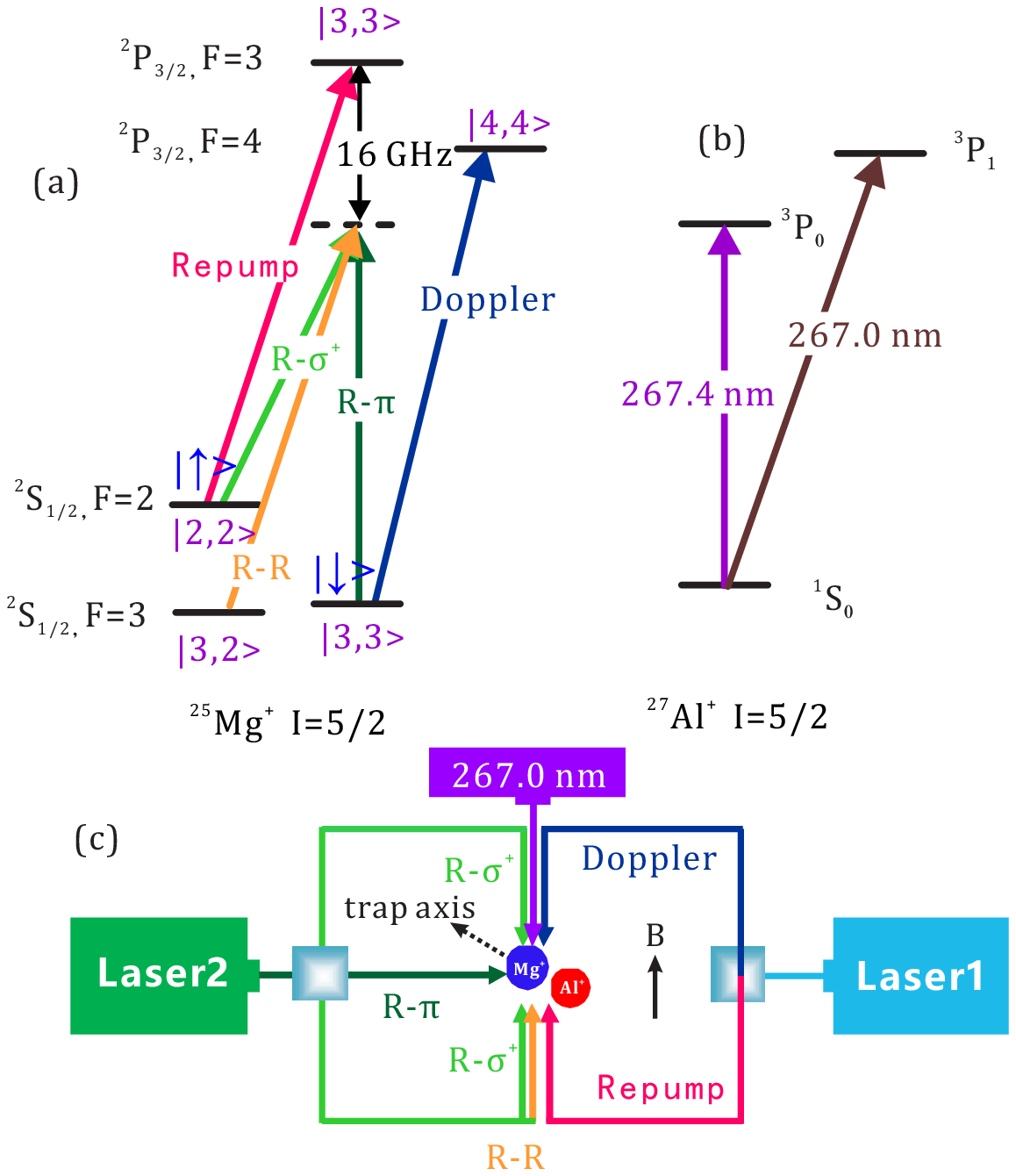}
 \caption{\label{EnergyLevel} (a) The energy level of  {$^{25}$Mg$^+$}. The states of {$^{25}$Mg$^+$} are labeled by the total angular momentum number $F$ and the magnetic quantum number $m_F$.  For {$^{25}$Mg$^+$}, $\mid\downarrow>$ $\equiv$ $^2$S$_{1/2}\mid3,3>$ and $\mid\uparrow>$ $\equiv$ $^2$S$_{1/2}\mid2,2>$. The Raman laser is red detuned 16 GHz, R-$\upsigma^+$ denotes Raman-$\upsigma^+$;  R-$\uppi$ denotes Raman-$\uppi$, R-R denotes Raman Repump.  (b) The energy level of {$^{27}$Al$^+$}.  The BSB and carrier transitions of {$^{27}$Al$^+$} are performed on $^1$S$_0 {\mid5/2, 5/2>}$ - $^3$P$_1{\mid7/2, 7/2>} $ transition at 267.0 nm. The clock transition of {$^{27}$Al$^+$} is $^1$S$_0$ - $^3$P$_0 $ at 267.4 nm. (c) The experimental setup. Two FHG lasers (Laser1 and Laser2) with a wavelength of 280 nm are used for Raman sideband cooling and Doppler cooling, respectively. The z direction is defined along the trap axis. The applied magnetic field is 6.5 G.}
 \end{figure}

As described previously \cite{Deng2015, Che2017, Xu2017}, a linear Paul trap is used to trap the {$^{25}$Mg$^+$} and {$^{27}$Al$^+$} ions pair. It consists of four blade electrodes and two end-cap tip electrodes. The blade electrodes supply the radial confinement of trapped ions and the end-cap  electrodes supply the axial confinement. The distance from the center of the trap to the blade electrodes is $r_0 = 0.8$ mm, and the radius of curvature of the blades is $r_e = 0.3$ mm and the collection solid angle is 0.8 sr \cite{Deng2015}. The distance between the center of the trap to the end-cap is $z = 3.0$ mm. The blade electrodes is driven by a helical resonator with a Q value over 400 and the end-cap is driven by a high voltage power supply directly. Three extra rods running along the trap axial direction are used for the compensation of stray electric fields in radial directions. The axial compensation is implemented on the end-cap electrodes.

Two re-entrant fused silica  viewports are installed on the vacuum chamber  to increase the collection efficiency of  the fluorescence laser and two sets of imaging lens are installed outside of the viewports for photon counting module and an electron multiplying CCD. The vacuum chamber with a pressure below $5 \times 10^{-9}$ Pa is maintained by a 40 L/s ion pump, nonevaporable getter pump and a titanium sublimation pump. Three pairs of Helmholtz coils powered by a precise current source  are mounted around the vacuum chamber to compensate background magnetic field  and  define the ion quantization axis with a magnetic field of 6.5 G. 

The {$^{25}$Mg$^+$} ion is loaded to the ion trap  
by two-photon ionization of  {$^{25}$Mg} atoms that are sputtered from a Mg wire through laser ablation. The ionization laser is a frequency-quadrupled 285 nm tunable diode laser system. 
The ablation laser is a commercially available passive Q-switched laser at a wavelength of 532 nm with a maximum pulse energy of 150 $\upmu$J, and a pulse duration of 2 ns. The repetition rate could be adjusted from 0 kHz to 11.0 kHz. The {$^{27}$Al$^+$} ion is loaded to the ion trap using the same laser ablation method. The ionization laser is an extended cavity diode laser (ECDL) with the frequency set to the resonance frequency of ${^2P_{3/2} (F'=4)} - {^2S_{1/2} (F=3)}$ transition which was measured to be 756.547133(3) THz \cite{Liu2018}. 

Two fourth-harmonic-generation (FHG)  lasers with a wavelength of 280 nm  are used for Raman sideband cooling and Doppler cooling of  the trapped  Mg ions, and one FHG laser with a wavelength of 267.0 nm is used for BSB and carrier QLS of {$^{27}$Al$^+$} ions. The laser frequencies of 280 nm are locked to a high precision wavemeter \cite{Zhang2013}. The one 280 nm FHG laser that is used for Doppler cooling, repumping and state detection is frequency shifted by two double passed AOM with a driving frequency of 450 MHz. The other 280 nm FHG laser is red-detuned 16 GHz from the resonance frequency, and is used as the Raman-$\uppi$ laser, the Raman-$\upsigma^{+}$ laser, and the Raman Repump laser. 
The master laser of the FHG at 267.0 nm is locked to a  high-finesse 10-cm long ultra-stable Fabry-P\'erot reference cavity and has  fractional frequency instability of $4.9 \times 10^{-16}$  which is limited by the thermal noise \cite{Zeng2018}.

\begin{figure}
 \includegraphics[width=8.5 cm]{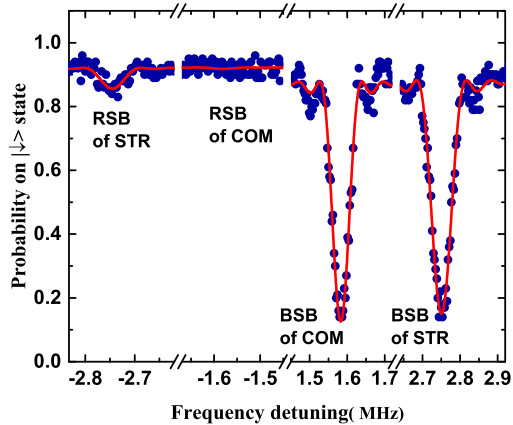}
 \caption{\label{COM} The Raman sideband spectroscopy of {$^{25}$Mg$^+$}. The secular motion frequency of STR mode is 2.75 MHz, and the secular motion frequency of COM mode is 1.58 MHz. The mean phonon numbers of the STR mode and the COM mode are  0.10(1) and  0.01(1), respectively.}
 \end{figure}

Figure \ref{EnergyLevel} shows the partial energy levels of Mg ion and Al ion. Doppler is the Doppler cooling beam and Repump is the repumping beam. Both the Doppler and Repump lasers are circularly polarized. R-$\upsigma^+$ and R-$\uppi$ are the two Raman laser beams used for the Raman sideband cooling. During the Raman sideband cooling process, the ${\mid 2,2>}$ state is pumped back to the ${\mid 3,3>} $ state by Repump laser. But the $^{25}$Mg$^+$ still has a small  probability to fall to the ${\mid3,2>}$ state. So the two Raman Repump beams, R-$\upsigma^+$  and R-R, are used to pump the ${\mid 3,2>}$ state to the ${\mid 2,2>}$ state. Here R-$\upsigma^+$  and R-R are circularly polarized ($\upsigma$) beam and R-$\uppi$ is linearly polarized ($\uppi$) beam. 

For Doppler cooling, the laser beam is red detuned by half of the linewidth with respect to the transition from $\mid\downarrow>$  to $^2$P$_ {3/2}\mid 4,4> $, which is the cycling transition for Doppler cooling and state detection. 
The frequency between $\mid\downarrow>$ and $\mid\uparrow>$ is measured by the RF resonant method \cite{Che2017} to implement Raman sideband  cooling. This frequency is also used to determine the magnetic field inside the vacuum chamber.
 
There are two motional modes in each direction which are named as ``stretch" (STR) mode and ``common" (COM) mode, respectively \cite{Chen2017, Cui2018}. Here we concentrate on the motional modes in axial direction of the ion trap for the demonstration of the QLS. Although it is  different from the single ions in ion traps, the cooling strategy of the ions pair can be optimized  based on the same strategy used for single ions. Based on our previous investigation on the cooling strategy of single ions \cite{Che2017}, we experimentally optimize the Raman sideband cooling strategy of the {$^{25}$Mg$^+$} and {$^{27}$Al$^+$}  ions pair for the QLS. For efficient cooling of the two motional modes, the second order  red-sideband (RSB) of STR mode, second order  RSB of COM mode, first order RSB of STR mode, and first order RSB of COM mode are carried out alternately. 

In the end, the motional modes of the {$^{25}$Mg$^+$} and {$^{27}$Al$^+$}  ions pair are cooled to the vibrational ground states, as shown in Fig.\ref{COM}. The mean phonon numbers and the heating rates are two important parameters in the QLS, which will determine the signal to noise ratio and the coherent time of the quantum logic process. The mean phonon number can be determined from the ratio of the RSB height and the BSB height. The COM mode is chosen as the transfer mode, and  the mean phonon number of the  COM mode is estimated to be $\bar{n}_{COM} = 0.01(1)$. Due to the sequence of the cooling strategy, the cooling effect for the STR mode is not as good as the COM mode, and the mean phonon number of STR mode is determined to be $\bar{n}_{STR} = 0.10(1)$. The heating rates of the two motional modes are measured by inserting a sequence of delay without any laser interactions after Raman sideband cooling and before the Raman sideband detection. The heating rate of the STR mode is measured to be 13(3) phonons/s, and the heating rate of the COM mode is measured to be 5.5(1) phonons/s. In comparison with the latest reports \cite{Chen2017, Cui2018}, the performance of resolved sideband cooling along the axial direction is good enough for the QLS.
The decoherence rate of the carrier oscillation and BSB oscillation are measured to be 21(4) $\upmu$s and 162(27) $\upmu$s, respectively.

\begin{figure}
 \includegraphics[width=8.5 cm]{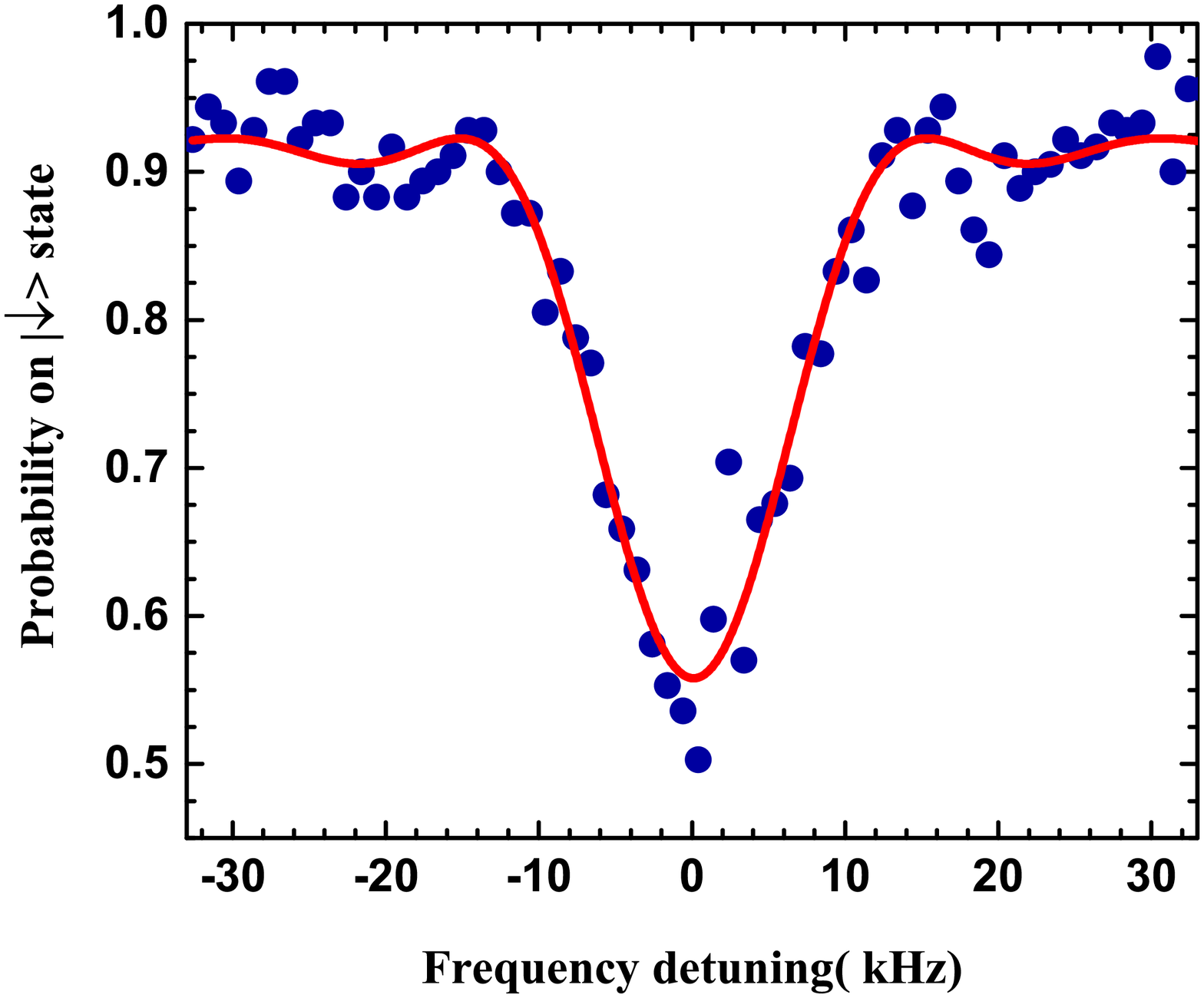}
 \caption{\label{QLS}The BSB spectroscopy of   {$^{27}$Al$^+$}   $^1$S$_0 {\mid5/2, 5/2>}$ - $^3$P$_1{\mid7/2, 7/2>} $  transition.}
 \end{figure}
 
After the ions are cooled to the vibrational ground states, the QLS can be carried out. A typical QLS sequence consists of following steps \cite{Schmidt2005}: (1) state preparation   so that the Al ion is prepared in the specific ground Zeeman state; (2) interrogation $\uppi$ pulse on the Al ion, which will change the internal state of the Al ion if the probing frequency is correct; (3) BSB  $\uppi$ pulse on the Al ion with the consequence to increase one phonon number of the ion pairs; (4) RSB Raman $\uppi$ pulse on the Mg ion to transfer the internal state of the Mg ion from $\mid\downarrow>$ to $\mid\uparrow>$; (5) detection pulses to determine the internal state of the Mg ion. When the Mg ion is on the $\mid\downarrow>$ state,  an average  photon number of 11 will be observed with 100 $\upmu$s detection time, and when it is on the $\mid\uparrow>$  the average photon number is about 1 with the same detection time \cite{Yuan2018}.
  
Before the carrier QLS detection, we scan the BSB frequency of the {$^{27}$Al$^+$}   $^1$S$_0 {\mid5/2, 5/2>}$ - $^3$P$_1{\mid7/2, 7/2>} $ transition without the interrogation pulses of {$^{27}$Al$^+$}.  Fig. \ref{QLS} shows the experimental data for the BSB spectroscopy. 
The BSB pulses of {$^{27}$Al$^+$} are applied with constant intensity and duration, corresponding to  $\uppi$ pulses on this transition when the laser beam is tuned to the resonance. Each data point is an average of  200 times. 
The observed linewidth of the transition is Fourier-transform limited with a fitted full width at half maximum (FWHM) of 13.5 kHz for an excitation pulse duration of t$_\pi$ = 60   $\upmu$s . The  contrast is limited by the fluctuation of the laser power and the detection laser frequency of  {$^{25}$Mg$^+$}, magnetic fluctuation of the system, and radial motion Debye-Waller factors.

After the BSB spectroscopy of {$^{27}$Al$^+$} is observed, the carrier QLS could be carried out by the {$^{27}$Al$^+$}   $^1$S$_0 {\mid5/2, 5/2>}$ - $^3$P$_1{\mid7/2, 7/2>} $ transition as an important step of the operation of {$^{27}$Al$^+$} ion optical clock. 
Fig. \ref{3P0} shows the Rabi spectroscopy and the Rabi oscillation on the transition of {$^{27}$Al$^+$}   $^1$S$_0 {\mid5/2, 5/2>}$ - $^3$P$_1{\mid7/2, 7/2>} $ based on QLS technology. The Rabi spectroscopy is carried out by varying the frequency of the interrogation  $\uppi$ pulses, and the BSB of  {$^{27}$Al$^+$} is set on resonance with $\uppi$ pulses. The FWHM of the Rabi spectroscopy is 77 kHz with a pulse duration of t$_\pi$ = 10   $\upmu$s. 

 \begin{figure}
 \includegraphics[width=8.5 cm]{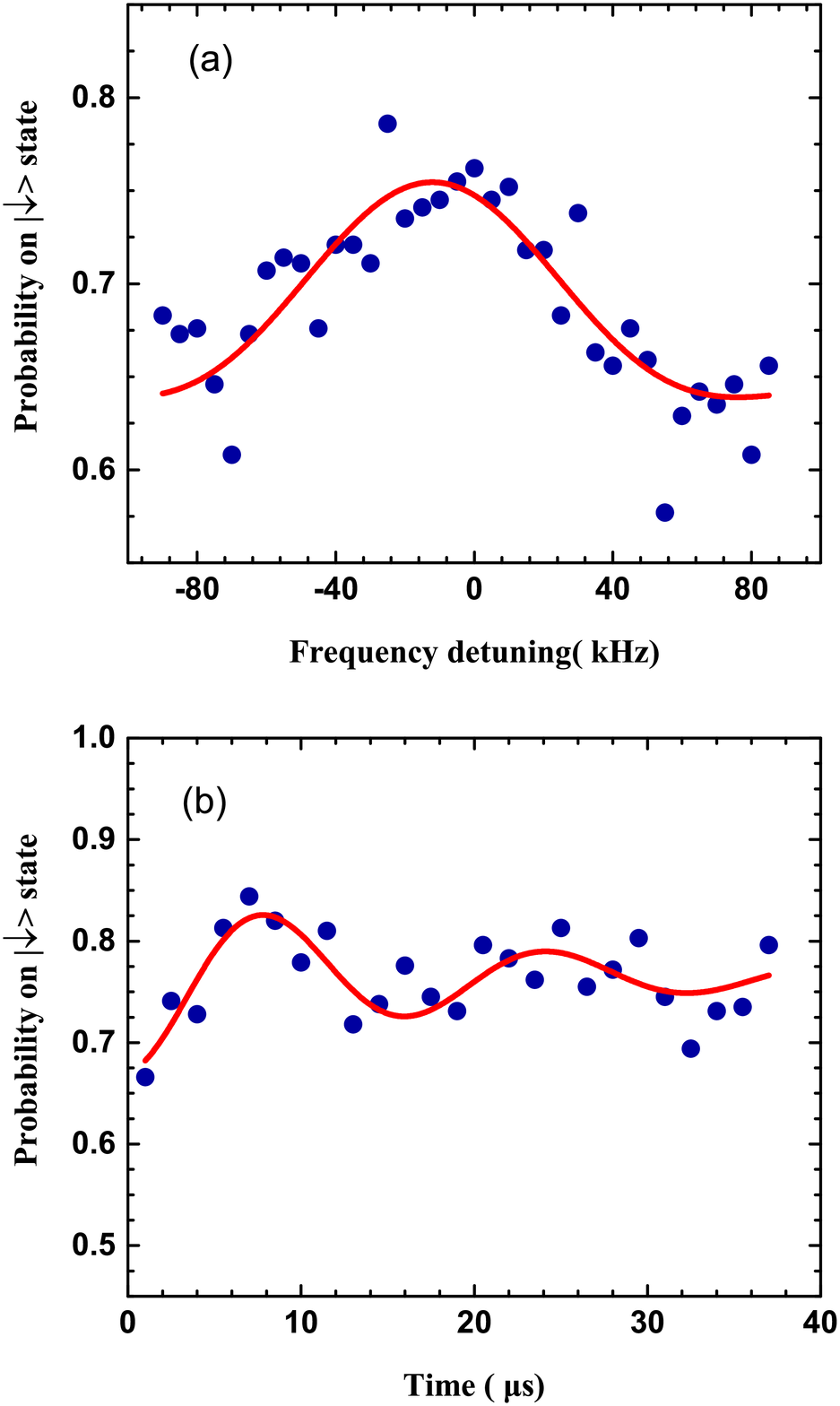}
 \caption{\label{3P0}(a) The {$^{27}$Al$^+$}  $^1$S$_0 {\mid5/2, 5/2>}$ - $^3$P$_1{\mid7/2, 7/2>} $  Rabi transition based on QLS. The data are fit by sinc$^2$ function. (b) Rabi oscillation observed at the center frequency of the Rabi transition. The data are fit by an exponentially damped sinusoidal function.}
 \end{figure}

The Rabi oscillation of internal state of  {$^{27}$Al$^+$} is obtained by varying the interrogation duration t$_i$  when the the frequency of interrogation pulses is set on the center of the Rabi spectroscopy. The observed coherent time of 18(8) $\upmu$s is limited by the  lifetime of the $^3$P$_1$ excited state, Rabi frequency fluctuations caused by the radial motion Debye-Waller factors, and magnetic field fluctuations.

In conclusion, we demonstrate the QLS for the $^1$S$_0 {\mid5/2, 5/2>}$ - $^3$P$_1{\mid7/2, 7/2>} $ transtion as a first step for the operation of Al$^+$ ion optical clock. As a premise for the QLS, we also realize Raman sideband cooling of the  {$^{27}$Al$^+$} with {$^{25}$Mg$^+$} ions pair to the motional ground states along the axial direction with the mean phonon number of 0.01(1) for the COM mode, and 0.10(1) for the STR mode. The  heating rate is evaluated to be 13(3) phonons/s for the STR mode and 5.5(1) phonons/s for the COM mode. Next we plan to Raman sideband cool the radial motional modes to the ground states, and to improve the contrast and coherent time of the {$^{27}$Al$^+$ $^1$S$_0$ - $^3$P$_1$ transition. With these improvements, we expect to detect the $^1$S$_0$ - $^3$P$_0$ clock transition through QLS in the near future.

\begin{acknowledgments}
We thank Dr. K. F. Cui for useful discussions. This work is partially supported by the National Key R$\&$D Program of China (Grant No. 2017YFA0304400) and the National Natural Science Foundation of China (Grants No. 11774108, No. 91336213 and No. 61875065).
\end{acknowledgments}

\bibliography{QLS}

\begin{thebibliography}{50}%
\makeatletter
\providecommand \@ifxundefined [1]{%
 \@ifx{#1\undefined}
}%
\providecommand \@ifnum [1]{%
 \ifnum #1\expandafter \@firstoftwo
 \else \expandafter \@secondoftwo
 \fi
}%
\providecommand \@ifx [1]{%
 \ifx #1\expandafter \@firstoftwo
 \else \expandafter \@secondoftwo
 \fi
}%
\providecommand \natexlab [1]{#1}%
\providecommand \enquote  [1]{``#1''}%
\providecommand \bibnamefont  [1]{#1}%
\providecommand \bibfnamefont [1]{#1}%
\providecommand \citenamefont [1]{#1}%
\providecommand \href@noop [0]{\@secondoftwo}%
\providecommand \href [0]{\begingroup \@sanitize@url \@href}%
\providecommand \@href[1]{\@@startlink{#1}\@@href}%
\providecommand \@@href[1]{\endgroup#1\@@endlink}%
\providecommand \@sanitize@url [0]{\catcode `\\12\catcode `\$12\catcode
  `\&12\catcode `\#12\catcode `\^12\catcode `\_12\catcode `\%12\relax}%
\providecommand \@@startlink[1]{}%
\providecommand \@@endlink[0]{}%
\providecommand \url  [0]{\begingroup\@sanitize@url \@url }%
\providecommand \@url [1]{\endgroup\@href {#1}{\urlprefix }}%
\providecommand \urlprefix  [0]{URL }%
\providecommand \Eprint [0]{\href }%
\providecommand \doibase [0]{http://dx.doi.org/}%
\providecommand \selectlanguage [0]{\@gobble}%
\providecommand \bibinfo  [0]{\@secondoftwo}%
\providecommand \bibfield  [0]{\@secondoftwo}%
\providecommand \translation [1]{[#1]}%
\providecommand \BibitemOpen [0]{}%
\providecommand \bibitemStop [0]{}%
\providecommand \bibitemNoStop [0]{.\EOS\space}%
\providecommand \EOS [0]{\spacefactor3000\relax}%
\providecommand \BibitemShut  [1]{\csname bibitem#1\endcsname}%
\let\auto@bib@innerbib\@empty
\bibitem [{\citenamefont {H\"ansch}(2006)}]{Haensch2006}%
  \BibitemOpen
  \bibfield  {author} {\bibinfo {author} {\bibfnamefont {T.~W.}\ \bibnamefont
  {H\"ansch}},\ }\href {\doibase 10.1103/revmodphys.78.1297} {\bibfield
  {journal} {\bibinfo  {journal} {Rev. Mod. Phys.}\ }\textbf {\bibinfo {volume}
  {78}},\ \bibinfo {pages} {1297} (\bibinfo {year} {2006})}\BibitemShut
  {NoStop}%
\bibitem [{\citenamefont {Chou}\ \emph {et~al.}(2010)\citenamefont {Chou},
  \citenamefont {Hume}, \citenamefont {Rosenband},\ and\ \citenamefont
  {Wineland}}]{Chou2010}%
  \BibitemOpen
  \bibfield  {author} {\bibinfo {author} {\bibfnamefont {C.~W.}\ \bibnamefont
  {Chou}}, \bibinfo {author} {\bibfnamefont {D.~B.}\ \bibnamefont {Hume}},
  \bibinfo {author} {\bibfnamefont {T.}~\bibnamefont {Rosenband}}, \ and\
  \bibinfo {author} {\bibfnamefont {D.~J.}\ \bibnamefont {Wineland}},\ }\href
  {\doibase 10.1126/science.1192720} {\bibfield  {journal} {\bibinfo  {journal}
  {Science}\ }\textbf {\bibinfo {volume} {329}},\ \bibinfo {pages} {1630}
  (\bibinfo {year} {2010})}\BibitemShut {NoStop}%
\bibitem [{\citenamefont {Steinmetz}\ \emph {et~al.}(2008)\citenamefont
  {Steinmetz}, \citenamefont {Wilken}, \citenamefont {Araujo-Hauck},
  \citenamefont {Holzwarth}, \citenamefont {H\"ansch}, \citenamefont
  {Pasquini}, \citenamefont {Manescau}, \citenamefont {D'Odorico},
  \citenamefont {Murphy}, \citenamefont {Kentischer}, \citenamefont {Schmidt},\
  and\ \citenamefont {Udem}}]{Steinmetz2008}%
  \BibitemOpen
  \bibfield  {author} {\bibinfo {author} {\bibfnamefont {T.}~\bibnamefont
  {Steinmetz}}, \bibinfo {author} {\bibfnamefont {T.}~\bibnamefont {Wilken}},
  \bibinfo {author} {\bibfnamefont {C.}~\bibnamefont {Araujo-Hauck}}, \bibinfo
  {author} {\bibfnamefont {R.}~\bibnamefont {Holzwarth}}, \bibinfo {author}
  {\bibfnamefont {T.~W.}\ \bibnamefont {H\"ansch}}, \bibinfo {author}
  {\bibfnamefont {L.}~\bibnamefont {Pasquini}}, \bibinfo {author}
  {\bibfnamefont {A.}~\bibnamefont {Manescau}}, \bibinfo {author}
  {\bibfnamefont {S.}~\bibnamefont {D'Odorico}}, \bibinfo {author}
  {\bibfnamefont {M.~T.}\ \bibnamefont {Murphy}}, \bibinfo {author}
  {\bibfnamefont {T.}~\bibnamefont {Kentischer}}, \bibinfo {author}
  {\bibfnamefont {W.}~\bibnamefont {Schmidt}}, \ and\ \bibinfo {author}
  {\bibfnamefont {T.}~\bibnamefont {Udem}},\ }\href {\doibase
  10.1126/science.1161030} {\bibfield  {journal} {\bibinfo  {journal}
  {Science}\ }\textbf {\bibinfo {volume} {321}},\ \bibinfo {pages} {1335}
  (\bibinfo {year} {2008})}\BibitemShut {NoStop}%
\bibitem [{\citenamefont {Godun}\ \emph {et~al.}(2014)\citenamefont {Godun},
  \citenamefont {Nisbet-Jones}, \citenamefont {Jones}, \citenamefont {King},
  \citenamefont {Johnson}, \citenamefont {Margolis}, \citenamefont {Szymaniec},
  \citenamefont {Lea}, \citenamefont {Bongs},\ and\ \citenamefont
  {Gill}}]{Godun2014}%
  \BibitemOpen
  \bibfield  {author} {\bibinfo {author} {\bibfnamefont {R.}~\bibnamefont
  {Godun}}, \bibinfo {author} {\bibfnamefont {P.}~\bibnamefont {Nisbet-Jones}},
  \bibinfo {author} {\bibfnamefont {J.}~\bibnamefont {Jones}}, \bibinfo
  {author} {\bibfnamefont {S.}~\bibnamefont {King}}, \bibinfo {author}
  {\bibfnamefont {L.}~\bibnamefont {Johnson}}, \bibinfo {author} {\bibfnamefont
  {H.}~\bibnamefont {Margolis}}, \bibinfo {author} {\bibfnamefont
  {K.}~\bibnamefont {Szymaniec}}, \bibinfo {author} {\bibfnamefont
  {S.}~\bibnamefont {Lea}}, \bibinfo {author} {\bibfnamefont {K.}~\bibnamefont
  {Bongs}}, \ and\ \bibinfo {author} {\bibfnamefont {P.}~\bibnamefont {Gill}},\
  }\href {\doibase 10.1103/physrevlett.113.210801} {\bibfield  {journal}
  {\bibinfo  {journal} {Phys. Rev. Lett.}\ }\textbf {\bibinfo {volume} {113}},\
  \bibinfo {pages} {210801} (\bibinfo {year} {2014})}\BibitemShut {NoStop}%
\bibitem [{\citenamefont {Huntemann}\ \emph {et~al.}(2014)\citenamefont
  {Huntemann}, \citenamefont {Lipphardt}, \citenamefont {Tamm}, \citenamefont
  {Gerginov}, \citenamefont {Weyers},\ and\ \citenamefont
  {Peik}}]{Huntemann2014}%
  \BibitemOpen
  \bibfield  {author} {\bibinfo {author} {\bibfnamefont {N.}~\bibnamefont
  {Huntemann}}, \bibinfo {author} {\bibfnamefont {B.}~\bibnamefont
  {Lipphardt}}, \bibinfo {author} {\bibfnamefont {C.}~\bibnamefont {Tamm}},
  \bibinfo {author} {\bibfnamefont {V.}~\bibnamefont {Gerginov}}, \bibinfo
  {author} {\bibfnamefont {S.}~\bibnamefont {Weyers}}, \ and\ \bibinfo {author}
  {\bibfnamefont {E.}~\bibnamefont {Peik}},\ }\href {\doibase
  10.1103/physrevlett.113.210802} {\bibfield  {journal} {\bibinfo  {journal}
  {Phys. Rev. Lett.}\ }\textbf {\bibinfo {volume} {113}},\ \bibinfo {pages}
  {090801} (\bibinfo {year} {2014})}\BibitemShut {NoStop}%
\bibitem [{\citenamefont {Bondarescu}\ \emph {et~al.}(2015)\citenamefont
  {Bondarescu}, \citenamefont {Schärer}, \citenamefont {Lundgren},
  \citenamefont {Het{\'{e}}nyi}, \citenamefont {Houli{\'{e}}}, \citenamefont
  {Jetzer},\ and\ \citenamefont {Bondarescu}}]{Bondarescu2015}%
  \BibitemOpen
  \bibfield  {author} {\bibinfo {author} {\bibfnamefont {R.}~\bibnamefont
  {Bondarescu}}, \bibinfo {author} {\bibfnamefont {A.}~\bibnamefont
  {Schärer}}, \bibinfo {author} {\bibfnamefont {A.}~\bibnamefont {Lundgren}},
  \bibinfo {author} {\bibfnamefont {G.}~\bibnamefont {Het{\'{e}}nyi}}, \bibinfo
  {author} {\bibfnamefont {N.}~\bibnamefont {Houli{\'{e}}}}, \bibinfo {author}
  {\bibfnamefont {P.}~\bibnamefont {Jetzer}}, \ and\ \bibinfo {author}
  {\bibfnamefont {M.}~\bibnamefont {Bondarescu}},\ }\href {\doibase
  10.1093/gji/ggv246} {\bibfield  {journal} {\bibinfo  {journal} {Geophys. J.
  Int.}\ }\textbf {\bibinfo {volume} {202}},\ \bibinfo {pages} {1770} (\bibinfo
  {year} {2015})}\BibitemShut {NoStop}%
\bibitem [{\citenamefont {Kolkowitz}\ \emph {et~al.}(2016)\citenamefont
  {Kolkowitz}, \citenamefont {Pikovski}, \citenamefont {Langellier},
  \citenamefont {Lukin}, \citenamefont {Walsworth},\ and\ \citenamefont
  {Ye}}]{Kolkowitz2016}%
  \BibitemOpen
  \bibfield  {author} {\bibinfo {author} {\bibfnamefont {S.}~\bibnamefont
  {Kolkowitz}}, \bibinfo {author} {\bibfnamefont {I.}~\bibnamefont {Pikovski}},
  \bibinfo {author} {\bibfnamefont {N.}~\bibnamefont {Langellier}}, \bibinfo
  {author} {\bibfnamefont {M.}~\bibnamefont {Lukin}}, \bibinfo {author}
  {\bibfnamefont {R.}~\bibnamefont {Walsworth}}, \ and\ \bibinfo {author}
  {\bibfnamefont {J.}~\bibnamefont {Ye}},\ }\href {\doibase
  10.1103/physrevd.94.124043} {\bibfield  {journal} {\bibinfo  {journal} {Phys.
  Rev. D}\ }\textbf {\bibinfo {volume} {94}},\ \bibinfo {pages} {124043}
  (\bibinfo {year} {2016})}\BibitemShut {NoStop}%
\bibitem [{\citenamefont {Delva}\ \emph {et~al.}(2017)\citenamefont {Delva},
  \citenamefont {Lodewyck}, \citenamefont {Bilicki}, \citenamefont {Bookjans},
  \citenamefont {Vallet}, \citenamefont {Targat}, \citenamefont {Pottie},
  \citenamefont {Guerlin}, \citenamefont {Meynadier}, \citenamefont
  {Poncin-Lafitte}, \citenamefont {Lopez}, \citenamefont {Amy-Klein},
  \citenamefont {Lee}, \citenamefont {Quintin}, \citenamefont {Lisdat},
  \citenamefont {Al-Masoudi}, \citenamefont {Dörscher}, \citenamefont
  {Grebing}, \citenamefont {Grosche}, \citenamefont {Kuhl}, \citenamefont
  {Raupach}, \citenamefont {Sterr}, \citenamefont {Hill}, \citenamefont
  {Hobson}, \citenamefont {Bowden}, \citenamefont {Kronjäger}, \citenamefont
  {Marra}, \citenamefont {Rolland}, \citenamefont {Baynes}, \citenamefont
  {Margolis},\ and\ \citenamefont {Gill}}]{Delva2017}%
  \BibitemOpen
  \bibfield  {author} {\bibinfo {author} {\bibfnamefont {P.}~\bibnamefont
  {Delva}}, \bibinfo {author} {\bibfnamefont {J.}~\bibnamefont {Lodewyck}},
  \bibinfo {author} {\bibfnamefont {S.}~\bibnamefont {Bilicki}}, \bibinfo
  {author} {\bibfnamefont {E.}~\bibnamefont {Bookjans}}, \bibinfo {author}
  {\bibfnamefont {G.}~\bibnamefont {Vallet}}, \bibinfo {author} {\bibfnamefont
  {R.~L.}\ \bibnamefont {Targat}}, \bibinfo {author} {\bibfnamefont {P.-E.}\
  \bibnamefont {Pottie}}, \bibinfo {author} {\bibfnamefont {C.}~\bibnamefont
  {Guerlin}}, \bibinfo {author} {\bibfnamefont {F.}~\bibnamefont {Meynadier}},
  \bibinfo {author} {\bibfnamefont {C.~L.}\ \bibnamefont {Poncin-Lafitte}},
  \bibinfo {author} {\bibfnamefont {O.}~\bibnamefont {Lopez}}, \bibinfo
  {author} {\bibfnamefont {A.}~\bibnamefont {Amy-Klein}}, \bibinfo {author}
  {\bibfnamefont {W.-K.}\ \bibnamefont {Lee}}, \bibinfo {author} {\bibfnamefont
  {N.}~\bibnamefont {Quintin}}, \bibinfo {author} {\bibfnamefont
  {C.}~\bibnamefont {Lisdat}}, \bibinfo {author} {\bibfnamefont
  {A.}~\bibnamefont {Al-Masoudi}}, \bibinfo {author} {\bibfnamefont
  {S.}~\bibnamefont {Dörscher}}, \bibinfo {author} {\bibfnamefont
  {C.}~\bibnamefont {Grebing}}, \bibinfo {author} {\bibfnamefont
  {G.}~\bibnamefont {Grosche}}, \bibinfo {author} {\bibfnamefont
  {A.}~\bibnamefont {Kuhl}}, \bibinfo {author} {\bibfnamefont {S.}~\bibnamefont
  {Raupach}}, \bibinfo {author} {\bibfnamefont {U.}~\bibnamefont {Sterr}},
  \bibinfo {author} {\bibfnamefont {I.}~\bibnamefont {Hill}}, \bibinfo {author}
  {\bibfnamefont {R.}~\bibnamefont {Hobson}}, \bibinfo {author} {\bibfnamefont
  {W.}~\bibnamefont {Bowden}}, \bibinfo {author} {\bibfnamefont
  {J.}~\bibnamefont {Kronjäger}}, \bibinfo {author} {\bibfnamefont
  {G.}~\bibnamefont {Marra}}, \bibinfo {author} {\bibfnamefont
  {A.}~\bibnamefont {Rolland}}, \bibinfo {author} {\bibfnamefont
  {F.}~\bibnamefont {Baynes}}, \bibinfo {author} {\bibfnamefont
  {H.}~\bibnamefont {Margolis}}, \ and\ \bibinfo {author} {\bibfnamefont
  {P.}~\bibnamefont {Gill}},\ }\href {\doibase 10.1103/physrevlett.118.221102}
  {\bibfield  {journal} {\bibinfo  {journal} {Phys. Rev. Lett.}\ }\textbf
  {\bibinfo {volume} {118}},\ \bibinfo {pages} {221102} (\bibinfo {year}
  {2017})}\BibitemShut {NoStop}%
\bibitem [{\citenamefont {Roberts}\ \emph {et~al.}(2017)\citenamefont
  {Roberts}, \citenamefont {Blewitt}, \citenamefont {Dailey}, \citenamefont
  {Murphy}, \citenamefont {Pospelov}, \citenamefont {Rollings}, \citenamefont
  {Sherman}, \citenamefont {Williams},\ and\ \citenamefont
  {Derevianko}}]{Roberts2017}%
  \BibitemOpen
  \bibfield  {author} {\bibinfo {author} {\bibfnamefont {B.~M.}\ \bibnamefont
  {Roberts}}, \bibinfo {author} {\bibfnamefont {G.}~\bibnamefont {Blewitt}},
  \bibinfo {author} {\bibfnamefont {C.}~\bibnamefont {Dailey}}, \bibinfo
  {author} {\bibfnamefont {M.}~\bibnamefont {Murphy}}, \bibinfo {author}
  {\bibfnamefont {M.}~\bibnamefont {Pospelov}}, \bibinfo {author}
  {\bibfnamefont {A.}~\bibnamefont {Rollings}}, \bibinfo {author}
  {\bibfnamefont {J.}~\bibnamefont {Sherman}}, \bibinfo {author} {\bibfnamefont
  {W.}~\bibnamefont {Williams}}, \ and\ \bibinfo {author} {\bibfnamefont
  {A.}~\bibnamefont {Derevianko}},\ }\href {\doibase
  10.1038/s41467-017-01440-4} {\bibfield  {journal} {\bibinfo  {journal} {Nat.
  Commun.}\ }\textbf {\bibinfo {volume} {8}},\ \bibinfo {pages} {1195}
  (\bibinfo {year} {2017})}\BibitemShut {NoStop}%
\bibitem [{\citenamefont {Campbell}\ \emph {et~al.}(2017)\citenamefont
  {Campbell}, \citenamefont {Hutson}, \citenamefont {Marti}, \citenamefont
  {Goban}, \citenamefont {Oppong}, \citenamefont {McNally}, \citenamefont
  {Sonderhouse}, \citenamefont {Robinson}, \citenamefont {Zhang}, \citenamefont
  {Bloom},\ and\ \citenamefont {Ye}}]{Campbell2017}%
  \BibitemOpen
  \bibfield  {author} {\bibinfo {author} {\bibfnamefont {S.~L.}\ \bibnamefont
  {Campbell}}, \bibinfo {author} {\bibfnamefont {R.~B.}\ \bibnamefont
  {Hutson}}, \bibinfo {author} {\bibfnamefont {G.~E.}\ \bibnamefont {Marti}},
  \bibinfo {author} {\bibfnamefont {A.}~\bibnamefont {Goban}}, \bibinfo
  {author} {\bibfnamefont {N.~D.}\ \bibnamefont {Oppong}}, \bibinfo {author}
  {\bibfnamefont {R.~L.}\ \bibnamefont {McNally}}, \bibinfo {author}
  {\bibfnamefont {L.}~\bibnamefont {Sonderhouse}}, \bibinfo {author}
  {\bibfnamefont {J.~M.}\ \bibnamefont {Robinson}}, \bibinfo {author}
  {\bibfnamefont {W.}~\bibnamefont {Zhang}}, \bibinfo {author} {\bibfnamefont
  {B.~J.}\ \bibnamefont {Bloom}}, \ and\ \bibinfo {author} {\bibfnamefont
  {J.}~\bibnamefont {Ye}},\ }\href {\doibase 10.1126/science.aam5538}
  {\bibfield  {journal} {\bibinfo  {journal} {Science}\ }\textbf {\bibinfo
  {volume} {358}},\ \bibinfo {pages} {90} (\bibinfo {year} {2017})}\BibitemShut
  {NoStop}%
\bibitem [{\citenamefont {McGrew}\ \emph {et~al.}(2018)\citenamefont {McGrew},
  \citenamefont {Zhang}, \citenamefont {Fasano}, \citenamefont {Schäffer},
  \citenamefont {Beloy}, \citenamefont {Nicolodi}, \citenamefont {Brown},
  \citenamefont {Hinkley}, \citenamefont {Milani}, \citenamefont {Schioppo},
  \citenamefont {Yoon},\ and\ \citenamefont {Ludlow}}]{McGrew2018}%
  \BibitemOpen
  \bibfield  {author} {\bibinfo {author} {\bibfnamefont {W.~F.}\ \bibnamefont
  {McGrew}}, \bibinfo {author} {\bibfnamefont {X.}~\bibnamefont {Zhang}},
  \bibinfo {author} {\bibfnamefont {R.~J.}\ \bibnamefont {Fasano}}, \bibinfo
  {author} {\bibfnamefont {S.~A.}\ \bibnamefont {Schäffer}}, \bibinfo {author}
  {\bibfnamefont {K.}~\bibnamefont {Beloy}}, \bibinfo {author} {\bibfnamefont
  {D.}~\bibnamefont {Nicolodi}}, \bibinfo {author} {\bibfnamefont {R.~C.}\
  \bibnamefont {Brown}}, \bibinfo {author} {\bibfnamefont {N.}~\bibnamefont
  {Hinkley}}, \bibinfo {author} {\bibfnamefont {G.}~\bibnamefont {Milani}},
  \bibinfo {author} {\bibfnamefont {M.}~\bibnamefont {Schioppo}}, \bibinfo
  {author} {\bibfnamefont {T.~H.}\ \bibnamefont {Yoon}}, \ and\ \bibinfo
  {author} {\bibfnamefont {A.~D.}\ \bibnamefont {Ludlow}},\ }\href {\doibase
  10.1038/s41586-018-0738-2} {\bibfield  {journal} {\bibinfo  {journal}
  {Nature}\ }\textbf {\bibinfo {volume} {564}},\ \bibinfo {pages} {87}
  (\bibinfo {year} {2018})}\BibitemShut {NoStop}%
\bibitem [{\citenamefont {Keller}\ \emph {et~al.}(2019)\citenamefont {Keller},
  \citenamefont {Burgermeister}, \citenamefont {Kalincev}, \citenamefont
  {Didier}, \citenamefont {Kulosa}, \citenamefont {Nordmann}, \citenamefont
  {Kiethe},\ and\ \citenamefont {Mehlstäubler}}]{Keller2019}%
  \BibitemOpen
  \bibfield  {author} {\bibinfo {author} {\bibfnamefont {J.}~\bibnamefont
  {Keller}}, \bibinfo {author} {\bibfnamefont {T.}~\bibnamefont
  {Burgermeister}}, \bibinfo {author} {\bibfnamefont {D.}~\bibnamefont
  {Kalincev}}, \bibinfo {author} {\bibfnamefont {A.}~\bibnamefont {Didier}},
  \bibinfo {author} {\bibfnamefont {A.~P.}\ \bibnamefont {Kulosa}}, \bibinfo
  {author} {\bibfnamefont {T.}~\bibnamefont {Nordmann}}, \bibinfo {author}
  {\bibfnamefont {J.}~\bibnamefont {Kiethe}}, \ and\ \bibinfo {author}
  {\bibfnamefont {T.~E.}\ \bibnamefont {Mehlstäubler}},\ }\href {\doibase
  10.1103/physreva.99.013405} {\bibfield  {journal} {\bibinfo  {journal} {Phys.
  Rev. A}\ }\textbf {\bibinfo {volume} {99}},\ \bibinfo {pages} {013405}
  (\bibinfo {year} {2019})}\BibitemShut {NoStop}%
\bibitem [{\citenamefont {Brewer}\ \emph {et~al.}()\citenamefont {Brewer},
  \citenamefont {Chen}, \citenamefont {Hankin}, \citenamefont {Clements},
  \citenamefont {Chou}, \citenamefont {Wineland}, \citenamefont {Hume},\ and\
  \citenamefont {Leibrandt}}]{Brewer2019}%
  \BibitemOpen
  \bibfield  {author} {\bibinfo {author} {\bibfnamefont {S.~M.}\ \bibnamefont
  {Brewer}}, \bibinfo {author} {\bibfnamefont {J.~S.}\ \bibnamefont {Chen}},
  \bibinfo {author} {\bibfnamefont {A.~M.}\ \bibnamefont {Hankin}}, \bibinfo
  {author} {\bibfnamefont {E.~R.}\ \bibnamefont {Clements}}, \bibinfo {author}
  {\bibfnamefont {C.~W.}\ \bibnamefont {Chou}}, \bibinfo {author}
  {\bibfnamefont {D.~J.}\ \bibnamefont {Wineland}}, \bibinfo {author}
  {\bibfnamefont {D.~B.}\ \bibnamefont {Hume}}, \ and\ \bibinfo {author}
  {\bibfnamefont {D.~R.}\ \bibnamefont {Leibrandt}},\ }\href@noop {} {\
  }\Eprint {http://arxiv.org/abs/http://arxiv.org/abs/1902.07694v1}
  {http://arxiv.org/abs/1902.07694v1} \BibitemShut {NoStop}%
\bibitem [{\citenamefont {Heavner}\ \emph {et~al.}(2014)\citenamefont
  {Heavner}, \citenamefont {Donley}, \citenamefont {Levi}, \citenamefont
  {Costanzo}, \citenamefont {Parker}, \citenamefont {Shirley}, \citenamefont
  {Ashby}, \citenamefont {Barlow},\ and\ \citenamefont
  {Jefferts}}]{Heavner2014}%
  \BibitemOpen
  \bibfield  {author} {\bibinfo {author} {\bibfnamefont {T.~P.}\ \bibnamefont
  {Heavner}}, \bibinfo {author} {\bibfnamefont {E.~A.}\ \bibnamefont {Donley}},
  \bibinfo {author} {\bibfnamefont {F.}~\bibnamefont {Levi}}, \bibinfo {author}
  {\bibfnamefont {G.}~\bibnamefont {Costanzo}}, \bibinfo {author}
  {\bibfnamefont {T.~E.}\ \bibnamefont {Parker}}, \bibinfo {author}
  {\bibfnamefont {J.~H.}\ \bibnamefont {Shirley}}, \bibinfo {author}
  {\bibfnamefont {N.}~\bibnamefont {Ashby}}, \bibinfo {author} {\bibfnamefont
  {S.}~\bibnamefont {Barlow}}, \ and\ \bibinfo {author} {\bibfnamefont {S.~R.}\
  \bibnamefont {Jefferts}},\ }\href {\doibase 10.1088/0026-1394/51/3/174}
  {\bibfield  {journal} {\bibinfo  {journal} {Metrologia}\ }\textbf {\bibinfo
  {volume} {51}},\ \bibinfo {pages} {174} (\bibinfo {year} {2014})}\BibitemShut
  {NoStop}%
\bibitem [{\citenamefont {Zhuang}\ \emph {et~al.}(2014)\citenamefont {Zhuang},
  \citenamefont {Zhang},\ and\ \citenamefont {Chen}}]{Zhuang2014}%
  \BibitemOpen
  \bibfield  {author} {\bibinfo {author} {\bibfnamefont {W.}~\bibnamefont
  {Zhuang}}, \bibinfo {author} {\bibfnamefont {T.-G.}\ \bibnamefont {Zhang}}, \
  and\ \bibinfo {author} {\bibfnamefont {J.-B.}\ \bibnamefont {Chen}},\ }\href
  {\doibase 10.1088/0256-307x/31/9/093201} {\bibfield  {journal} {\bibinfo
  {journal} {Chin. Phys. Lett.}\ }\textbf {\bibinfo {volume} {31}},\ \bibinfo
  {pages} {093201} (\bibinfo {year} {2014})}\BibitemShut {NoStop}%
\bibitem [{\citenamefont {Oskay}\ \emph {et~al.}(2006)\citenamefont {Oskay},
  \citenamefont {Diddams}, \citenamefont {Donley}, \citenamefont {Fortier},
  \citenamefont {Heavner}, \citenamefont {Hollberg}, \citenamefont {Itano},
  \citenamefont {Jefferts}, \citenamefont {Delaney}, \citenamefont {Kim},
  \citenamefont {Levi}, \citenamefont {Parker},\ and\ \citenamefont
  {Bergquist}}]{Oskay2006}%
  \BibitemOpen
  \bibfield  {author} {\bibinfo {author} {\bibfnamefont {W.~H.}\ \bibnamefont
  {Oskay}}, \bibinfo {author} {\bibfnamefont {S.~A.}\ \bibnamefont {Diddams}},
  \bibinfo {author} {\bibfnamefont {E.~A.}\ \bibnamefont {Donley}}, \bibinfo
  {author} {\bibfnamefont {T.~M.}\ \bibnamefont {Fortier}}, \bibinfo {author}
  {\bibfnamefont {T.~P.}\ \bibnamefont {Heavner}}, \bibinfo {author}
  {\bibfnamefont {L.}~\bibnamefont {Hollberg}}, \bibinfo {author}
  {\bibfnamefont {W.~M.}\ \bibnamefont {Itano}}, \bibinfo {author}
  {\bibfnamefont {S.~R.}\ \bibnamefont {Jefferts}}, \bibinfo {author}
  {\bibfnamefont {M.~J.}\ \bibnamefont {Delaney}}, \bibinfo {author}
  {\bibfnamefont {K.}~\bibnamefont {Kim}}, \bibinfo {author} {\bibfnamefont
  {F.}~\bibnamefont {Levi}}, \bibinfo {author} {\bibfnamefont {T.~E.}\
  \bibnamefont {Parker}}, \ and\ \bibinfo {author} {\bibfnamefont {J.~C.}\
  \bibnamefont {Bergquist}},\ }\href {\doibase 10.1103/physrevlett.97.020801}
  {\bibfield  {journal} {\bibinfo  {journal} {Phys. Rev. Lett.}\ }\textbf
  {\bibinfo {volume} {97}},\ \bibinfo {pages} {020801} (\bibinfo {year}
  {2006})}\BibitemShut {NoStop}%
\bibitem [{\citenamefont {Liu}\ \emph {et~al.}(2019)\citenamefont {Liu},
  \citenamefont {Zou}, \citenamefont {He}, \citenamefont {Chen}, \citenamefont
  {Shen},\ and\ \citenamefont {Yuan}}]{Liu2019}%
  \BibitemOpen
  \bibfield  {author} {\bibinfo {author} {\bibfnamefont {Q.}~\bibnamefont
  {Liu}}, \bibinfo {author} {\bibfnamefont {H.}~\bibnamefont {Zou}}, \bibinfo
  {author} {\bibfnamefont {X.}~\bibnamefont {He}}, \bibinfo {author}
  {\bibfnamefont {G.}~\bibnamefont {Chen}}, \bibinfo {author} {\bibfnamefont
  {Y.}~\bibnamefont {Shen}}, \ and\ \bibinfo {author} {\bibfnamefont
  {J.}~\bibnamefont {Yuan}},\ }\href {\doibase 10.1063/1.5068692} {\bibfield
  {journal} {\bibinfo  {journal} {Rev. Sci. Instrum.}\ }\textbf {\bibinfo
  {volume} {90}},\ \bibinfo {pages} {013107} (\bibinfo {year}
  {2019})}\BibitemShut {NoStop}%
\bibitem [{\citenamefont {Barwood}\ \emph {et~al.}(2014)\citenamefont
  {Barwood}, \citenamefont {Huang}, \citenamefont {Klein}, \citenamefont
  {Johnson}, \citenamefont {King}, \citenamefont {Margolis}, \citenamefont
  {Szymaniec},\ and\ \citenamefont {Gill}}]{Barwood2014}%
  \BibitemOpen
  \bibfield  {author} {\bibinfo {author} {\bibfnamefont {G.~P.}\ \bibnamefont
  {Barwood}}, \bibinfo {author} {\bibfnamefont {G.}~\bibnamefont {Huang}},
  \bibinfo {author} {\bibfnamefont {H.~A.}\ \bibnamefont {Klein}}, \bibinfo
  {author} {\bibfnamefont {L.~A.~M.}\ \bibnamefont {Johnson}}, \bibinfo
  {author} {\bibfnamefont {S.~A.}\ \bibnamefont {King}}, \bibinfo {author}
  {\bibfnamefont {H.~S.}\ \bibnamefont {Margolis}}, \bibinfo {author}
  {\bibfnamefont {K.}~\bibnamefont {Szymaniec}}, \ and\ \bibinfo {author}
  {\bibfnamefont {P.}~\bibnamefont {Gill}},\ }\href {\doibase
  10.1103/physreva.89.050501} {\bibfield  {journal} {\bibinfo  {journal} {Phys.
  Rev. A}\ }\textbf {\bibinfo {volume} {89}},\ \bibinfo {pages} {050501}
  (\bibinfo {year} {2014})}\BibitemShut {NoStop}%
\bibitem [{\citenamefont {Dub{\'{e}}}\ \emph {et~al.}(2014)\citenamefont
  {Dub{\'{e}}}, \citenamefont {Madej}, \citenamefont {Tibbo},\ and\
  \citenamefont {Bernard}}]{Dube2014}%
  \BibitemOpen
  \bibfield  {author} {\bibinfo {author} {\bibfnamefont {P.}~\bibnamefont
  {Dub{\'{e}}}}, \bibinfo {author} {\bibfnamefont {A.~A.}\ \bibnamefont
  {Madej}}, \bibinfo {author} {\bibfnamefont {M.}~\bibnamefont {Tibbo}}, \ and\
  \bibinfo {author} {\bibfnamefont {J.~E.}\ \bibnamefont {Bernard}},\ }\href
  {\doibase 10.1103/physrevlett.112.173002} {\bibfield  {journal} {\bibinfo
  {journal} {Phys. Rev. Lett.}\ }\textbf {\bibinfo {volume} {112}},\ \bibinfo
  {pages} {020801} (\bibinfo {year} {2014})}\BibitemShut {NoStop}%
\bibitem [{\citenamefont {King}\ \emph {et~al.}(2012)\citenamefont {King},
  \citenamefont {Godun}, \citenamefont {Webster}, \citenamefont {Margolis},
  \citenamefont {Johnson}, \citenamefont {Szymaniec}, \citenamefont {Baird},\
  and\ \citenamefont {Gill}}]{King2012}%
  \BibitemOpen
  \bibfield  {author} {\bibinfo {author} {\bibfnamefont {S.~A.}\ \bibnamefont
  {King}}, \bibinfo {author} {\bibfnamefont {R.~M.}\ \bibnamefont {Godun}},
  \bibinfo {author} {\bibfnamefont {S.~A.}\ \bibnamefont {Webster}}, \bibinfo
  {author} {\bibfnamefont {H.~S.}\ \bibnamefont {Margolis}}, \bibinfo {author}
  {\bibfnamefont {L.~A.~M.}\ \bibnamefont {Johnson}}, \bibinfo {author}
  {\bibfnamefont {K.}~\bibnamefont {Szymaniec}}, \bibinfo {author}
  {\bibfnamefont {P.~E.~G.}\ \bibnamefont {Baird}}, \ and\ \bibinfo {author}
  {\bibfnamefont {P.}~\bibnamefont {Gill}},\ }\href {\doibase
  10.1088/1367-2630/14/1/013045} {\bibfield  {journal} {\bibinfo  {journal}
  {New J. Phys.}\ }\textbf {\bibinfo {volume} {14}},\ \bibinfo {pages} {013045}
  (\bibinfo {year} {2012})}\BibitemShut {NoStop}%
\bibitem [{\citenamefont {Huntemann}\ \emph {et~al.}(2012)\citenamefont
  {Huntemann}, \citenamefont {Okhapkin}, \citenamefont {Lipphardt},
  \citenamefont {Weyers}, \citenamefont {Tamm},\ and\ \citenamefont
  {Peik}}]{Huntemann2012}%
  \BibitemOpen
  \bibfield  {author} {\bibinfo {author} {\bibfnamefont {N.}~\bibnamefont
  {Huntemann}}, \bibinfo {author} {\bibfnamefont {M.}~\bibnamefont {Okhapkin}},
  \bibinfo {author} {\bibfnamefont {B.}~\bibnamefont {Lipphardt}}, \bibinfo
  {author} {\bibfnamefont {S.}~\bibnamefont {Weyers}}, \bibinfo {author}
  {\bibfnamefont {C.}~\bibnamefont {Tamm}}, \ and\ \bibinfo {author}
  {\bibfnamefont {E.}~\bibnamefont {Peik}},\ }\href {\doibase
  10.1103/physrevlett.108.090801} {\bibfield  {journal} {\bibinfo  {journal}
  {Phys. Rev. Lett.}\ }\textbf {\bibinfo {volume} {108}},\ \bibinfo {pages}
  {090801} (\bibinfo {year} {2012})}\BibitemShut {NoStop}%
\bibitem [{\citenamefont {Hashimoto}\ \emph {et~al.}(2011)\citenamefont
  {Hashimoto}, \citenamefont {Kitaoka}, \citenamefont {Yoshida},\ and\
  \citenamefont {Hasegawa}}]{Hashimoto2011}%
  \BibitemOpen
  \bibfield  {author} {\bibinfo {author} {\bibfnamefont {Y.}~\bibnamefont
  {Hashimoto}}, \bibinfo {author} {\bibfnamefont {M.}~\bibnamefont {Kitaoka}},
  \bibinfo {author} {\bibfnamefont {T.}~\bibnamefont {Yoshida}}, \ and\
  \bibinfo {author} {\bibfnamefont {S.}~\bibnamefont {Hasegawa}},\ }\href
  {\doibase 10.1007/s00340-011-4471-x} {\bibfield  {journal} {\bibinfo
  {journal} {Appl. Phys. B}\ }\textbf {\bibinfo {volume} {103}},\ \bibinfo
  {pages} {339} (\bibinfo {year} {2011})}\BibitemShut {NoStop}%
\bibitem [{\citenamefont {Liu}\ \emph {et~al.}(2014)\citenamefont {Liu},
  \citenamefont {Huang}, \citenamefont {Bian}, \citenamefont {Shao},
  \citenamefont {Qian}, \citenamefont {Guan},\ and\ \citenamefont
  {Gao}}]{Liu2014}%
  \BibitemOpen
  \bibfield  {author} {\bibinfo {author} {\bibfnamefont {P.-L.}\ \bibnamefont
  {Liu}}, \bibinfo {author} {\bibfnamefont {Y.}~\bibnamefont {Huang}}, \bibinfo
  {author} {\bibfnamefont {W.}~\bibnamefont {Bian}}, \bibinfo {author}
  {\bibfnamefont {H.}~\bibnamefont {Shao}}, \bibinfo {author} {\bibfnamefont
  {Y.}~\bibnamefont {Qian}}, \bibinfo {author} {\bibfnamefont {H.}~\bibnamefont
  {Guan}}, \ and\ \bibinfo {author} {\bibfnamefont {K.-L.}\ \bibnamefont
  {Gao}},\ }\href {\doibase 10.1088/0256-307x/31/11/113702} {\bibfield
  {journal} {\bibinfo  {journal} {Chin. Phys. Lett.}\ }\textbf {\bibinfo
  {volume} {31}},\ \bibinfo {pages} {113702} (\bibinfo {year}
  {2014})}\BibitemShut {NoStop}%
\bibitem [{\citenamefont {Wang}\ \emph {et~al.}(2007)\citenamefont {Wang},
  \citenamefont {Dumke}, \citenamefont {Liu}, \citenamefont {Stejskal},
  \citenamefont {Zhao}, \citenamefont {Zhang}, \citenamefont {Lu},
  \citenamefont {Wang}, \citenamefont {Becker},\ and\ \citenamefont
  {Walther}}]{Wang2007}%
  \BibitemOpen
  \bibfield  {author} {\bibinfo {author} {\bibfnamefont {Y.}~\bibnamefont
  {Wang}}, \bibinfo {author} {\bibfnamefont {R.}~\bibnamefont {Dumke}},
  \bibinfo {author} {\bibfnamefont {T.}~\bibnamefont {Liu}}, \bibinfo {author}
  {\bibfnamefont {A.}~\bibnamefont {Stejskal}}, \bibinfo {author}
  {\bibfnamefont {Y.}~\bibnamefont {Zhao}}, \bibinfo {author} {\bibfnamefont
  {J.}~\bibnamefont {Zhang}}, \bibinfo {author} {\bibfnamefont
  {Z.}~\bibnamefont {Lu}}, \bibinfo {author} {\bibfnamefont {L.}~\bibnamefont
  {Wang}}, \bibinfo {author} {\bibfnamefont {T.}~\bibnamefont {Becker}}, \ and\
  \bibinfo {author} {\bibfnamefont {H.}~\bibnamefont {Walther}},\ }\href
  {\doibase 10.1016/j.optcom.2007.01.068} {\bibfield  {journal} {\bibinfo
  {journal} {Opt. Commun.}\ }\textbf {\bibinfo {volume} {273}},\ \bibinfo
  {pages} {526} (\bibinfo {year} {2007})}\BibitemShut {NoStop}%
\bibitem [{\citenamefont {Koerber}\ \emph {et~al.}(2002)\citenamefont
  {Koerber}, \citenamefont {Schacht}, \citenamefont {Hendrickson},
  \citenamefont {Nagourney},\ and\ \citenamefont {Fortson}}]{Koerber2002}%
  \BibitemOpen
  \bibfield  {author} {\bibinfo {author} {\bibfnamefont {T.~W.}\ \bibnamefont
  {Koerber}}, \bibinfo {author} {\bibfnamefont {M.~H.}\ \bibnamefont
  {Schacht}}, \bibinfo {author} {\bibfnamefont {K.~R.~G.}\ \bibnamefont
  {Hendrickson}}, \bibinfo {author} {\bibfnamefont {W.}~\bibnamefont
  {Nagourney}}, \ and\ \bibinfo {author} {\bibfnamefont {E.~N.}\ \bibnamefont
  {Fortson}},\ }\href {\doibase 10.1103/physrevlett.88.143002} {\bibfield
  {journal} {\bibinfo  {journal} {Phys. Rev. Lett.}\ }\textbf {\bibinfo
  {volume} {88}},\ \bibinfo {pages} {143002} (\bibinfo {year}
  {2002})}\BibitemShut {NoStop}%
\bibitem [{\citenamefont {Takamoto}\ \emph {et~al.}(2011)\citenamefont
  {Takamoto}, \citenamefont {Takano},\ and\ \citenamefont
  {Katori}}]{Takamoto2011}%
  \BibitemOpen
  \bibfield  {author} {\bibinfo {author} {\bibfnamefont {M.}~\bibnamefont
  {Takamoto}}, \bibinfo {author} {\bibfnamefont {T.}~\bibnamefont {Takano}}, \
  and\ \bibinfo {author} {\bibfnamefont {H.}~\bibnamefont {Katori}},\ }\href
  {\doibase 10.1038/nphoton.2011.34} {\bibfield  {journal} {\bibinfo  {journal}
  {Nat. Photon.}\ }\textbf {\bibinfo {volume} {5}},\ \bibinfo {pages} {288}
  (\bibinfo {year} {2011})}\BibitemShut {NoStop}%
\bibitem [{\citenamefont {Bloom}\ \emph {et~al.}(2014)\citenamefont {Bloom},
  \citenamefont {Nicholson}, \citenamefont {Williams}, \citenamefont
  {Campbell}, \citenamefont {Bishof}, \citenamefont {Zhang}, \citenamefont
  {Zhang}, \citenamefont {Bromley},\ and\ \citenamefont {Ye}}]{Bloom2014}%
  \BibitemOpen
  \bibfield  {author} {\bibinfo {author} {\bibfnamefont {B.~J.}\ \bibnamefont
  {Bloom}}, \bibinfo {author} {\bibfnamefont {T.~L.}\ \bibnamefont
  {Nicholson}}, \bibinfo {author} {\bibfnamefont {J.~R.}\ \bibnamefont
  {Williams}}, \bibinfo {author} {\bibfnamefont {S.~L.}\ \bibnamefont
  {Campbell}}, \bibinfo {author} {\bibfnamefont {M.}~\bibnamefont {Bishof}},
  \bibinfo {author} {\bibfnamefont {X.}~\bibnamefont {Zhang}}, \bibinfo
  {author} {\bibfnamefont {W.}~\bibnamefont {Zhang}}, \bibinfo {author}
  {\bibfnamefont {S.~L.}\ \bibnamefont {Bromley}}, \ and\ \bibinfo {author}
  {\bibfnamefont {J.}~\bibnamefont {Ye}},\ }\href {\doibase
  10.1038/nature12941} {\bibfield  {journal} {\bibinfo  {journal} {Nature}\
  }\textbf {\bibinfo {volume} {506}},\ \bibinfo {pages} {71} (\bibinfo {year}
  {2014})}\BibitemShut {NoStop}%
\bibitem [{\citenamefont {Lin}\ \emph {et~al.}(2015)\citenamefont {Lin},
  \citenamefont {Wang}, \citenamefont {Li}, \citenamefont {Meng}, \citenamefont
  {Lin}, \citenamefont {Zang}, \citenamefont {Sun}, \citenamefont {Fang},
  \citenamefont {Li},\ and\ \citenamefont {Fang}}]{Lin2015}%
  \BibitemOpen
  \bibfield  {author} {\bibinfo {author} {\bibfnamefont {Y.-G.}\ \bibnamefont
  {Lin}}, \bibinfo {author} {\bibfnamefont {Q.}~\bibnamefont {Wang}}, \bibinfo
  {author} {\bibfnamefont {Y.}~\bibnamefont {Li}}, \bibinfo {author}
  {\bibfnamefont {F.}~\bibnamefont {Meng}}, \bibinfo {author} {\bibfnamefont
  {B.-K.}\ \bibnamefont {Lin}}, \bibinfo {author} {\bibfnamefont {E.-J.}\
  \bibnamefont {Zang}}, \bibinfo {author} {\bibfnamefont {Z.}~\bibnamefont
  {Sun}}, \bibinfo {author} {\bibfnamefont {F.}~\bibnamefont {Fang}}, \bibinfo
  {author} {\bibfnamefont {T.-C.}\ \bibnamefont {Li}}, \ and\ \bibinfo {author}
  {\bibfnamefont {Z.-J.}\ \bibnamefont {Fang}},\ }\href {\doibase
  10.1088/0256-307x/32/9/090601} {\bibfield  {journal} {\bibinfo  {journal}
  {Chin. Phys. Lett.}\ }\textbf {\bibinfo {volume} {32}},\ \bibinfo {pages}
  {090601} (\bibinfo {year} {2015})}\BibitemShut {NoStop}%
\bibitem [{\citenamefont {Wang}\ \emph {et~al.}(2018)\citenamefont {Wang},
  \citenamefont {Yin}, \citenamefont {Ren}, \citenamefont {Xu}, \citenamefont
  {Lu}, \citenamefont {Han}, \citenamefont {Guo},\ and\ \citenamefont
  {Chang}}]{Wang2018}%
  \BibitemOpen
  \bibfield  {author} {\bibinfo {author} {\bibfnamefont {Y.-B.}\ \bibnamefont
  {Wang}}, \bibinfo {author} {\bibfnamefont {M.-J.}\ \bibnamefont {Yin}},
  \bibinfo {author} {\bibfnamefont {J.}~\bibnamefont {Ren}}, \bibinfo {author}
  {\bibfnamefont {Q.-F.}\ \bibnamefont {Xu}}, \bibinfo {author} {\bibfnamefont
  {B.-Q.}\ \bibnamefont {Lu}}, \bibinfo {author} {\bibfnamefont {J.-X.}\
  \bibnamefont {Han}}, \bibinfo {author} {\bibfnamefont {Y.}~\bibnamefont
  {Guo}}, \ and\ \bibinfo {author} {\bibfnamefont {H.}~\bibnamefont {Chang}},\
  }\href {\doibase 10.1088/1674-1056/27/2/023701} {\bibfield  {journal}
  {\bibinfo  {journal} {Chin. Phys. B}\ }\textbf {\bibinfo {volume} {27}},\
  \bibinfo {pages} {023701} (\bibinfo {year} {2018})}\BibitemShut {NoStop}%
\bibitem [{\citenamefont {Hinkley}\ \emph {et~al.}(2013)\citenamefont
  {Hinkley}, \citenamefont {Sherman}, \citenamefont {Phillips}, \citenamefont
  {Schioppo}, \citenamefont {Lemke}, \citenamefont {Beloy}, \citenamefont
  {Pizzocaro}, \citenamefont {Oates},\ and\ \citenamefont
  {Ludlow}}]{Hinkley2013}%
  \BibitemOpen
  \bibfield  {author} {\bibinfo {author} {\bibfnamefont {N.}~\bibnamefont
  {Hinkley}}, \bibinfo {author} {\bibfnamefont {J.~A.}\ \bibnamefont
  {Sherman}}, \bibinfo {author} {\bibfnamefont {N.~B.}\ \bibnamefont
  {Phillips}}, \bibinfo {author} {\bibfnamefont {M.}~\bibnamefont {Schioppo}},
  \bibinfo {author} {\bibfnamefont {N.~D.}\ \bibnamefont {Lemke}}, \bibinfo
  {author} {\bibfnamefont {K.}~\bibnamefont {Beloy}}, \bibinfo {author}
  {\bibfnamefont {M.}~\bibnamefont {Pizzocaro}}, \bibinfo {author}
  {\bibfnamefont {C.~W.}\ \bibnamefont {Oates}}, \ and\ \bibinfo {author}
  {\bibfnamefont {A.~D.}\ \bibnamefont {Ludlow}},\ }\href {\doibase
  10.1126/science.1240420} {\bibfield  {journal} {\bibinfo  {journal}
  {Science}\ }\textbf {\bibinfo {volume} {341}},\ \bibinfo {pages} {1215}
  (\bibinfo {year} {2013})}\BibitemShut {NoStop}%
\bibitem [{\citenamefont {Zhang}\ \emph {et~al.}(2016)\citenamefont {Zhang},
  \citenamefont {Liu}, \citenamefont {Zhang}, \citenamefont {Jiang},
  \citenamefont {Xiong}, \citenamefont {Lü},\ and\ \citenamefont
  {He}}]{Zhang2016}%
  \BibitemOpen
  \bibfield  {author} {\bibinfo {author} {\bibfnamefont {M.-J.}\ \bibnamefont
  {Zhang}}, \bibinfo {author} {\bibfnamefont {H.}~\bibnamefont {Liu}}, \bibinfo
  {author} {\bibfnamefont {X.}~\bibnamefont {Zhang}}, \bibinfo {author}
  {\bibfnamefont {K.-L.}\ \bibnamefont {Jiang}}, \bibinfo {author}
  {\bibfnamefont {Z.-X.}\ \bibnamefont {Xiong}}, \bibinfo {author}
  {\bibfnamefont {B.-L.}\ \bibnamefont {Lü}}, \ and\ \bibinfo {author}
  {\bibfnamefont {L.-X.}\ \bibnamefont {He}},\ }\href {\doibase
  10.1088/0256-307x/33/7/070601} {\bibfield  {journal} {\bibinfo  {journal}
  {Chin. Phys. Lett.}\ }\textbf {\bibinfo {volume} {33}},\ \bibinfo {pages}
  {070601} (\bibinfo {year} {2016})}\BibitemShut {NoStop}%
\bibitem [{\citenamefont {McFerran}\ \emph {et~al.}(2012)\citenamefont
  {McFerran}, \citenamefont {Yi}, \citenamefont {Mejri}, \citenamefont {Manno},
  \citenamefont {Zhang}, \citenamefont {Gu{\'{e}}na}, \citenamefont {Coq},\
  and\ \citenamefont {Bize}}]{McFerran2012}%
  \BibitemOpen
  \bibfield  {author} {\bibinfo {author} {\bibfnamefont {J.~J.}\ \bibnamefont
  {McFerran}}, \bibinfo {author} {\bibfnamefont {L.}~\bibnamefont {Yi}},
  \bibinfo {author} {\bibfnamefont {S.}~\bibnamefont {Mejri}}, \bibinfo
  {author} {\bibfnamefont {S.~D.}\ \bibnamefont {Manno}}, \bibinfo {author}
  {\bibfnamefont {W.}~\bibnamefont {Zhang}}, \bibinfo {author} {\bibfnamefont
  {J.}~\bibnamefont {Gu{\'{e}}na}}, \bibinfo {author} {\bibfnamefont {Y.~L.}\
  \bibnamefont {Coq}}, \ and\ \bibinfo {author} {\bibfnamefont
  {S.}~\bibnamefont {Bize}},\ }\href {\doibase 10.1103/physrevlett.108.183004}
  {\bibfield  {journal} {\bibinfo  {journal} {Phys. Rev. Lett.}\ }\textbf
  {\bibinfo {volume} {108}},\ \bibinfo {pages} {183004} (\bibinfo {year}
  {2012})}\BibitemShut {NoStop}%
\bibitem [{\citenamefont {Liu}\ \emph {et~al.}(2013)\citenamefont {Liu},
  \citenamefont {Yin}, \citenamefont {Liu}, \citenamefont {Qian}, \citenamefont
  {Xu}, \citenamefont {Hong},\ and\ \citenamefont {Wang}}]{Liu2013}%
  \BibitemOpen
  \bibfield  {author} {\bibinfo {author} {\bibfnamefont {H.-L.}\ \bibnamefont
  {Liu}}, \bibinfo {author} {\bibfnamefont {S.-Q.}\ \bibnamefont {Yin}},
  \bibinfo {author} {\bibfnamefont {K.-K.}\ \bibnamefont {Liu}}, \bibinfo
  {author} {\bibfnamefont {J.}~\bibnamefont {Qian}}, \bibinfo {author}
  {\bibfnamefont {Z.}~\bibnamefont {Xu}}, \bibinfo {author} {\bibfnamefont
  {T.}~\bibnamefont {Hong}}, \ and\ \bibinfo {author} {\bibfnamefont {Y.-Z.}\
  \bibnamefont {Wang}},\ }\href {\doibase 10.1088/1674-1056/22/4/043701}
  {\bibfield  {journal} {\bibinfo  {journal} {Chin. Phys. B}\ }\textbf
  {\bibinfo {volume} {22}},\ \bibinfo {pages} {043701} (\bibinfo {year}
  {2013})}\BibitemShut {NoStop}%
\bibitem [{\citenamefont {Yu}\ \emph {et~al.}(1992)\citenamefont {Yu},
  \citenamefont {Dehmelt},\ and\ \citenamefont {Nagourney}}]{Yu1992}%
  \BibitemOpen
  \bibfield  {author} {\bibinfo {author} {\bibfnamefont {N.}~\bibnamefont
  {Yu}}, \bibinfo {author} {\bibfnamefont {H.}~\bibnamefont {Dehmelt}}, \ and\
  \bibinfo {author} {\bibfnamefont {W.}~\bibnamefont {Nagourney}},\ }\href
  {\doibase 10.1073/pnas.89.16.7289} {\bibfield  {journal} {\bibinfo  {journal}
  {Proc. Natl. Acad. Sci. USA}\ }\textbf {\bibinfo {volume} {89}},\ \bibinfo
  {pages} {7289} (\bibinfo {year} {1992})}\BibitemShut {NoStop}%
\bibitem [{\citenamefont {Safronova}\ \emph {et~al.}(2011)\citenamefont
  {Safronova}, \citenamefont {Kozlov},\ and\ \citenamefont
  {Clark}}]{Safronova2011}%
  \BibitemOpen
  \bibfield  {author} {\bibinfo {author} {\bibfnamefont {M.~S.}\ \bibnamefont
  {Safronova}}, \bibinfo {author} {\bibfnamefont {M.~G.}\ \bibnamefont
  {Kozlov}}, \ and\ \bibinfo {author} {\bibfnamefont {C.~W.}\ \bibnamefont
  {Clark}},\ }\href {\doibase 10.1103/physrevlett.107.143006} {\bibfield
  {journal} {\bibinfo  {journal} {Phys. Rev. Lett.}\ }\textbf {\bibinfo
  {volume} {107}},\ \bibinfo {pages} {143006} (\bibinfo {year}
  {2011})}\BibitemShut {NoStop}%
\bibitem [{\citenamefont {Zhang}\ \emph {et~al.}(2017)\citenamefont {Zhang},
  \citenamefont {Deng}, \citenamefont {Luo},\ and\ \citenamefont
  {Lu}}]{Zhang2017}%
  \BibitemOpen
  \bibfield  {author} {\bibinfo {author} {\bibfnamefont {J.}~\bibnamefont
  {Zhang}}, \bibinfo {author} {\bibfnamefont {K.}~\bibnamefont {Deng}},
  \bibinfo {author} {\bibfnamefont {J.}~\bibnamefont {Luo}}, \ and\ \bibinfo
  {author} {\bibfnamefont {Z.-H.}\ \bibnamefont {Lu}},\ }\href {\doibase
  10.1088/0256-307x/34/5/050601} {\bibfield  {journal} {\bibinfo  {journal}
  {Chin. Phys. Lett.}\ }\textbf {\bibinfo {volume} {34}},\ \bibinfo {pages}
  {050601} (\bibinfo {year} {2017})}\BibitemShut {NoStop}%
\bibitem [{\citenamefont {Schmidt}\ \emph {et~al.}(2005)\citenamefont
  {Schmidt}, \citenamefont {Rosenband}, \citenamefont {Lange}, \citenamefont
  {Itano}, \citenamefont {Bergquist},\ and\ \citenamefont
  {Wineland}}]{Schmidt2005}%
  \BibitemOpen
  \bibfield  {author} {\bibinfo {author} {\bibfnamefont {P.~O.}\ \bibnamefont
  {Schmidt}}, \bibinfo {author} {\bibfnamefont {T.}~\bibnamefont {Rosenband}},
  \bibinfo {author} {\bibfnamefont {C.}~\bibnamefont {Lange}}, \bibinfo
  {author} {\bibfnamefont {W.}~\bibnamefont {Itano}}, \bibinfo {author}
  {\bibfnamefont {J.}~\bibnamefont {Bergquist}}, \ and\ \bibinfo {author}
  {\bibfnamefont {D.~J.}\ \bibnamefont {Wineland}},\ }\href {\doibase
  10.1126/science.1114375} {\bibfield  {journal} {\bibinfo  {journal}
  {Science}\ }\textbf {\bibinfo {volume} {309}},\ \bibinfo {pages} {749}
  (\bibinfo {year} {2005})}\BibitemShut {NoStop}%
\bibitem [{\citenamefont {Rosenband}\ \emph {et~al.}(2007)\citenamefont
  {Rosenband}, \citenamefont {Schmidt}, \citenamefont {Hume}, \citenamefont
  {Itano}, \citenamefont {Fortier}, \citenamefont {Stalnaker}, \citenamefont
  {Kim}, \citenamefont {Diddams}, \citenamefont {Koelemeij}, \citenamefont
  {Bergquist},\ and\ \citenamefont {Wineland}}]{Rosenband2007}%
  \BibitemOpen
  \bibfield  {author} {\bibinfo {author} {\bibfnamefont {T.}~\bibnamefont
  {Rosenband}}, \bibinfo {author} {\bibfnamefont {P.}~\bibnamefont {Schmidt}},
  \bibinfo {author} {\bibfnamefont {D.}~\bibnamefont {Hume}}, \bibinfo {author}
  {\bibfnamefont {W.}~\bibnamefont {Itano}}, \bibinfo {author} {\bibfnamefont
  {T.}~\bibnamefont {Fortier}}, \bibinfo {author} {\bibfnamefont
  {J.}~\bibnamefont {Stalnaker}}, \bibinfo {author} {\bibfnamefont
  {K.}~\bibnamefont {Kim}}, \bibinfo {author} {\bibfnamefont {S.}~\bibnamefont
  {Diddams}}, \bibinfo {author} {\bibfnamefont {J.}~\bibnamefont {Koelemeij}},
  \bibinfo {author} {\bibfnamefont {J.}~\bibnamefont {Bergquist}}, \ and\
  \bibinfo {author} {\bibfnamefont {D.}~\bibnamefont {Wineland}},\ }\href
  {\doibase 10.1103/physrevlett.98.220801} {\bibfield  {journal} {\bibinfo
  {journal} {Phys. Rev. Lett.}\ }\textbf {\bibinfo {volume} {98}},\ \bibinfo
  {pages} {220801} (\bibinfo {year} {2007})}\BibitemShut {NoStop}%
\bibitem [{\citenamefont {Guggemos}\ \emph {et~al.}(2015)\citenamefont
  {Guggemos}, \citenamefont {Heinrich}, \citenamefont {Herrera-Sancho},
  \citenamefont {Blatt},\ and\ \citenamefont {Roos}}]{Guggemos2015}%
  \BibitemOpen
  \bibfield  {author} {\bibinfo {author} {\bibfnamefont {M.}~\bibnamefont
  {Guggemos}}, \bibinfo {author} {\bibfnamefont {D.}~\bibnamefont {Heinrich}},
  \bibinfo {author} {\bibfnamefont {O.~A.}\ \bibnamefont {Herrera-Sancho}},
  \bibinfo {author} {\bibfnamefont {R.}~\bibnamefont {Blatt}}, \ and\ \bibinfo
  {author} {\bibfnamefont {C.~F.}\ \bibnamefont {Roos}},\ }\href {\doibase
  10.1088/1367-2630/17/10/103001} {\bibfield  {journal} {\bibinfo  {journal}
  {New J. Phys.}\ }\textbf {\bibinfo {volume} {17}},\ \bibinfo {pages} {103001}
  (\bibinfo {year} {2015})}\BibitemShut {NoStop}%
\bibitem [{\citenamefont {Cui}\ \emph {et~al.}(2018)\citenamefont {Cui},
  \citenamefont {Shang}, \citenamefont {Chao}, \citenamefont {Wang},
  \citenamefont {Yuan}, \citenamefont {Zhang}, \citenamefont {Cao},
  \citenamefont {Shu},\ and\ \citenamefont {Huang}}]{Cui2018}%
  \BibitemOpen
  \bibfield  {author} {\bibinfo {author} {\bibfnamefont {K.-F.}\ \bibnamefont
  {Cui}}, \bibinfo {author} {\bibfnamefont {J.-J.}\ \bibnamefont {Shang}},
  \bibinfo {author} {\bibfnamefont {S.-J.}\ \bibnamefont {Chao}}, \bibinfo
  {author} {\bibfnamefont {S.-M.}\ \bibnamefont {Wang}}, \bibinfo {author}
  {\bibfnamefont {J.-B.}\ \bibnamefont {Yuan}}, \bibinfo {author}
  {\bibfnamefont {P.}~\bibnamefont {Zhang}}, \bibinfo {author} {\bibfnamefont
  {J.}~\bibnamefont {Cao}}, \bibinfo {author} {\bibfnamefont {H.-L.}\
  \bibnamefont {Shu}}, \ and\ \bibinfo {author} {\bibfnamefont {X.-R.}\
  \bibnamefont {Huang}},\ }\href {\doibase 10.1088/1361-6455/aaa591} {\bibfield
   {journal} {\bibinfo  {journal} {J. Phys. B: At., Mol. Opt. Phys.}\ }\textbf
  {\bibinfo {volume} {51}},\ \bibinfo {pages} {045502} (\bibinfo {year}
  {2018})}\BibitemShut {NoStop}%
\bibitem [{\citenamefont {Ksenia}\ \emph {et~al.}(2018)\citenamefont {Ksenia},
  \citenamefont {Ilia}, \citenamefont {Ilya}, \citenamefont {Alexander},\ and\
  \citenamefont {Nikolay}}]{Ksenia2018}%
  \BibitemOpen
  \bibfield  {author} {\bibinfo {author} {\bibfnamefont {K.}~\bibnamefont
  {Ksenia}}, \bibinfo {author} {\bibfnamefont {Z.}~\bibnamefont {Ilia}},
  \bibinfo {author} {\bibfnamefont {S.}~\bibnamefont {Ilya}}, \bibinfo {author}
  {\bibfnamefont {B.}~\bibnamefont {Alexander}}, \ and\ \bibinfo {author}
  {\bibfnamefont {K.}~\bibnamefont {Nikolay}},\ }in\ \href {\doibase
  10.1109/eftf.2018.8409073} {\emph {\bibinfo {booktitle} {2018 European
  Frequency and Time Forum ({EFTF})}}}\ (\bibinfo  {publisher} {{IEEE}},\
  \bibinfo {year} {2018})\BibitemShut {NoStop}%
\bibitem [{\citenamefont {Hannig}\ \emph {et~al.}(2019)\citenamefont {Hannig},
  \citenamefont {Pelzer}, \citenamefont {Scharnhorst}, \citenamefont {Kramer},
  \citenamefont {Stepanova}, \citenamefont {Xu}, \citenamefont {Spethmann},
  \citenamefont {Leroux}, \citenamefont {Mehlstäubler},\ and\ \citenamefont
  {Schmidt}}]{Hannig2019}%
  \BibitemOpen
  \bibfield  {author} {\bibinfo {author} {\bibfnamefont {S.}~\bibnamefont
  {Hannig}}, \bibinfo {author} {\bibfnamefont {L.}~\bibnamefont {Pelzer}},
  \bibinfo {author} {\bibfnamefont {N.}~\bibnamefont {Scharnhorst}}, \bibinfo
  {author} {\bibfnamefont {J.}~\bibnamefont {Kramer}}, \bibinfo {author}
  {\bibfnamefont {M.}~\bibnamefont {Stepanova}}, \bibinfo {author}
  {\bibfnamefont {Z.~T.}\ \bibnamefont {Xu}}, \bibinfo {author} {\bibfnamefont
  {N.}~\bibnamefont {Spethmann}}, \bibinfo {author} {\bibfnamefont {I.~D.}\
  \bibnamefont {Leroux}}, \bibinfo {author} {\bibfnamefont {T.~E.}\
  \bibnamefont {Mehlstäubler}}, \ and\ \bibinfo {author} {\bibfnamefont
  {P.~O.}\ \bibnamefont {Schmidt}},\ }\href@noop {} {\bibfield  {journal}
  {\bibinfo  {journal} {arXiv}\ } (\bibinfo {year} {2019})},\ \Eprint
  {http://arxiv.org/abs/http://arxiv.org/abs/1901.02250v1}
  {http://arxiv.org/abs/1901.02250v1} \BibitemShut {NoStop}%
\bibitem [{\citenamefont {Deng}\ \emph {et~al.}(2015)\citenamefont {Deng},
  \citenamefont {Che}, \citenamefont {Lan}, \citenamefont {Ge}, \citenamefont
  {Xu}, \citenamefont {Yuan}, \citenamefont {Zhang},\ and\ \citenamefont
  {Lu}}]{Deng2015}%
  \BibitemOpen
  \bibfield  {author} {\bibinfo {author} {\bibfnamefont {K.}~\bibnamefont
  {Deng}}, \bibinfo {author} {\bibfnamefont {H.}~\bibnamefont {Che}}, \bibinfo
  {author} {\bibfnamefont {Y.}~\bibnamefont {Lan}}, \bibinfo {author}
  {\bibfnamefont {Y.~P.}\ \bibnamefont {Ge}}, \bibinfo {author} {\bibfnamefont
  {Z.~T.}\ \bibnamefont {Xu}}, \bibinfo {author} {\bibfnamefont {W.~H.}\
  \bibnamefont {Yuan}}, \bibinfo {author} {\bibfnamefont {J.}~\bibnamefont
  {Zhang}}, \ and\ \bibinfo {author} {\bibfnamefont {Z.~H.}\ \bibnamefont
  {Lu}},\ }\href {\doibase 10.1063/1.4931420} {\bibfield  {journal} {\bibinfo
  {journal} {J. Appl. Phys.}\ }\textbf {\bibinfo {volume} {118}},\ \bibinfo
  {pages} {113106} (\bibinfo {year} {2015})}\BibitemShut {NoStop}%
\bibitem [{\citenamefont {Che}\ \emph {et~al.}(2017)\citenamefont {Che},
  \citenamefont {Deng}, \citenamefont {Xu}, \citenamefont {Yuan}, \citenamefont
  {Zhang},\ and\ \citenamefont {Lu}}]{Che2017}%
  \BibitemOpen
  \bibfield  {author} {\bibinfo {author} {\bibfnamefont {H.}~\bibnamefont
  {Che}}, \bibinfo {author} {\bibfnamefont {K.}~\bibnamefont {Deng}}, \bibinfo
  {author} {\bibfnamefont {Z.~T.}\ \bibnamefont {Xu}}, \bibinfo {author}
  {\bibfnamefont {W.~H.}\ \bibnamefont {Yuan}}, \bibinfo {author}
  {\bibfnamefont {J.}~\bibnamefont {Zhang}}, \ and\ \bibinfo {author}
  {\bibfnamefont {Z.~H.}\ \bibnamefont {Lu}},\ }\href {\doibase
  10.1103/physreva.96.013417} {\bibfield  {journal} {\bibinfo  {journal} {Phys.
  Rev. A}\ }\textbf {\bibinfo {volume} {96}},\ \bibinfo {pages} {013417}
  (\bibinfo {year} {2017})}\BibitemShut {NoStop}%
\bibitem [{\citenamefont {Xu}\ \emph {et~al.}(2017)\citenamefont {Xu},
  \citenamefont {Deng}, \citenamefont {Che}, \citenamefont {Yuan},
  \citenamefont {Zhang},\ and\ \citenamefont {Lu}}]{Xu2017}%
  \BibitemOpen
  \bibfield  {author} {\bibinfo {author} {\bibfnamefont {Z.~T.}\ \bibnamefont
  {Xu}}, \bibinfo {author} {\bibfnamefont {K.}~\bibnamefont {Deng}}, \bibinfo
  {author} {\bibfnamefont {H.}~\bibnamefont {Che}}, \bibinfo {author}
  {\bibfnamefont {W.~H.}\ \bibnamefont {Yuan}}, \bibinfo {author}
  {\bibfnamefont {J.}~\bibnamefont {Zhang}}, \ and\ \bibinfo {author}
  {\bibfnamefont {Z.~H.}\ \bibnamefont {Lu}},\ }\href {\doibase
  10.1103/physreva.96.052507} {\bibfield  {journal} {\bibinfo  {journal} {Phys.
  Rev. A}\ }\textbf {\bibinfo {volume} {96}},\ \bibinfo {pages} {052507}
  (\bibinfo {year} {2017})}\BibitemShut {NoStop}%
\bibitem [{\citenamefont {Liu}\ \emph {et~al.}(2018)\citenamefont {Liu},
  \citenamefont {Yuan}, \citenamefont {Cheng}, \citenamefont {Wang},
  \citenamefont {Xu}, \citenamefont {Deng},\ and\ \citenamefont
  {Lu}}]{Liu2018}%
  \BibitemOpen
  \bibfield  {author} {\bibinfo {author} {\bibfnamefont {H.}~\bibnamefont
  {Liu}}, \bibinfo {author} {\bibfnamefont {W.}~\bibnamefont {Yuan}}, \bibinfo
  {author} {\bibfnamefont {F.}~\bibnamefont {Cheng}}, \bibinfo {author}
  {\bibfnamefont {Z.}~\bibnamefont {Wang}}, \bibinfo {author} {\bibfnamefont
  {Z.}~\bibnamefont {Xu}}, \bibinfo {author} {\bibfnamefont {K.}~\bibnamefont
  {Deng}}, \ and\ \bibinfo {author} {\bibfnamefont {Z.}~\bibnamefont {Lu}},\
  }\href {\doibase 10.1088/1361-6455/aada9a} {\bibfield  {journal} {\bibinfo
  {journal} {J. Phys. B: At., Mol. Opt. Phys.}\ }\textbf {\bibinfo {volume}
  {51}},\ \bibinfo {pages} {225002} (\bibinfo {year} {2018})}\BibitemShut
  {NoStop}%
\bibitem [{\citenamefont {Zhang}\ \emph {et~al.}(2013)\citenamefont {Zhang},
  \citenamefont {Yuan}, \citenamefont {Deng}, \citenamefont {Deng},
  \citenamefont {Xu}, \citenamefont {Qin}, \citenamefont {Lu},\ and\
  \citenamefont {Luo}}]{Zhang2013}%
  \BibitemOpen
  \bibfield  {author} {\bibinfo {author} {\bibfnamefont {J.}~\bibnamefont
  {Zhang}}, \bibinfo {author} {\bibfnamefont {W.~H.}\ \bibnamefont {Yuan}},
  \bibinfo {author} {\bibfnamefont {K.}~\bibnamefont {Deng}}, \bibinfo {author}
  {\bibfnamefont {A.}~\bibnamefont {Deng}}, \bibinfo {author} {\bibfnamefont
  {Z.~T.}\ \bibnamefont {Xu}}, \bibinfo {author} {\bibfnamefont {C.~B.}\
  \bibnamefont {Qin}}, \bibinfo {author} {\bibfnamefont {Z.~H.}\ \bibnamefont
  {Lu}}, \ and\ \bibinfo {author} {\bibfnamefont {J.}~\bibnamefont {Luo}},\
  }\href {\doibase 10.1063/1.4847135} {\bibfield  {journal} {\bibinfo
  {journal} {Rev. Sci. Instrum.}\ }\textbf {\bibinfo {volume} {84}},\ \bibinfo
  {pages} {123109} (\bibinfo {year} {2013})}\BibitemShut {NoStop}%
\bibitem [{\citenamefont {Zeng}\ \emph {et~al.}(2018)\citenamefont {Zeng},
  \citenamefont {Ye}, \citenamefont {Shi}, \citenamefont {Wang}, \citenamefont
  {Deng}, \citenamefont {Zhang},\ and\ \citenamefont {Lu}}]{Zeng2018}%
  \BibitemOpen
  \bibfield  {author} {\bibinfo {author} {\bibfnamefont {X.~Y.}\ \bibnamefont
  {Zeng}}, \bibinfo {author} {\bibfnamefont {Y.~X.}\ \bibnamefont {Ye}},
  \bibinfo {author} {\bibfnamefont {X.~H.}\ \bibnamefont {Shi}}, \bibinfo
  {author} {\bibfnamefont {Z.~Y.}\ \bibnamefont {Wang}}, \bibinfo {author}
  {\bibfnamefont {K.}~\bibnamefont {Deng}}, \bibinfo {author} {\bibfnamefont
  {J.}~\bibnamefont {Zhang}}, \ and\ \bibinfo {author} {\bibfnamefont {Z.~H.}\
  \bibnamefont {Lu}},\ }\href {\doibase 10.1364/ol.43.001690} {\bibfield
  {journal} {\bibinfo  {journal} {Opt. Lett.}\ }\textbf {\bibinfo {volume}
  {43}},\ \bibinfo {pages} {1690} (\bibinfo {year} {2018})}\BibitemShut
  {NoStop}%
\bibitem [{\citenamefont {Chen}\ \emph {et~al.}(2017)\citenamefont {Chen},
  \citenamefont {Brewer}, \citenamefont {Chou}, \citenamefont {Wineland},
  \citenamefont {Leibrandt},\ and\ \citenamefont {Hume}}]{Chen2017}%
  \BibitemOpen
  \bibfield  {author} {\bibinfo {author} {\bibfnamefont {J.-S.}\ \bibnamefont
  {Chen}}, \bibinfo {author} {\bibfnamefont {S.}~\bibnamefont {Brewer}},
  \bibinfo {author} {\bibfnamefont {C.}~\bibnamefont {Chou}}, \bibinfo {author}
  {\bibfnamefont {D.}~\bibnamefont {Wineland}}, \bibinfo {author}
  {\bibfnamefont {D.}~\bibnamefont {Leibrandt}}, \ and\ \bibinfo {author}
  {\bibfnamefont {D.}~\bibnamefont {Hume}},\ }\href {\doibase
  10.1103/physrevlett.118.053002} {\bibfield  {journal} {\bibinfo  {journal}
  {Phys. Rev. Lett.}\ }\textbf {\bibinfo {volume} {118}},\ \bibinfo {pages}
  {053002} (\bibinfo {year} {2017})}\BibitemShut {NoStop}%
\bibitem [{\citenamefont {Yuan}\ \emph {et~al.}(2018)\citenamefont {Yuan},
  \citenamefont {Deng}, \citenamefont {Ma}, \citenamefont {Che}, \citenamefont
  {Xu}, \citenamefont {Liu}, \citenamefont {Zhang},\ and\ \citenamefont
  {Lu}}]{Yuan2018}%
  \BibitemOpen
  \bibfield  {author} {\bibinfo {author} {\bibfnamefont {W.~H.}\ \bibnamefont
  {Yuan}}, \bibinfo {author} {\bibfnamefont {K.}~\bibnamefont {Deng}}, \bibinfo
  {author} {\bibfnamefont {Z.~Y.}\ \bibnamefont {Ma}}, \bibinfo {author}
  {\bibfnamefont {H.}~\bibnamefont {Che}}, \bibinfo {author} {\bibfnamefont
  {Z.~T.}\ \bibnamefont {Xu}}, \bibinfo {author} {\bibfnamefont {H.~L.}\
  \bibnamefont {Liu}}, \bibinfo {author} {\bibfnamefont {J.}~\bibnamefont
  {Zhang}}, \ and\ \bibinfo {author} {\bibfnamefont {Z.~H.}\ \bibnamefont
  {Lu}},\ }\href {\doibase 10.1103/physreva.98.052507} {\bibfield  {journal}
  {\bibinfo  {journal} {Phys. Rev. A}\ }\textbf {\bibinfo {volume} {98}},\
  \bibinfo {pages} {052507} (\bibinfo {year} {2018})}\BibitemShut {NoStop}%
\end{thebibliography}%

\end{document}